\pdfoutput=1 

\documentclass[aps,pra,showpacs,twocolumn,amsmath,amssymb,superscriptaddress,footinbib]{revtex4}
\usepackage{mathptmx}
\usepackage{subfigure}
\usepackage{dcolumn}
\usepackage{amsmath,amssymb}
\usepackage{bm}
\usepackage{color}
\usepackage{overpic}
\usepackage{latexsym}
\usepackage{epstopdf}
\usepackage{color}
\usepackage[english]{babel}
\usepackage{latexsym}

\usepackage{graphicx} 
\usepackage{epsf} 
\usepackage{subfigure} 
\usepackage{amsmath} 
\usepackage{amssymb} 
\usepackage{amsfonts}
\usepackage{bm}
\usepackage{natbib}
\usepackage{epstopdf}
\usepackage{appendix}

\definecolor{mygrey}{gray}{0.35}
\definecolor{mygreen}{rgb}{0.85,1,0.9}
\definecolor{myzard}{cmyk}{0,0,0.05,0}
\definecolor{mywhite}{rgb}{1,1,1}
\definecolor{myred}{rgb}{1,0,0}

\usepackage[colorlinks=true,citecolor=blue,linkcolor=blue]{hyperref}

\def\be{\begin{equation}}
\def\ee{\end{equation}}
\def\ba{\begin{align}}
\def\enda{\end{align}}
\def\bi{\begin{itemize}}
\def\ei{\end{itemize}}

\def\txt#1{\textrm{#1}}

 \def\ee{\mathord{\rm e}}
 
 \def\ii{\mathord{\rm i}}

\def\half{\textstyle\frac{1}{2}}

\renewcommand{\ii}{{\rm i}}
\renewcommand{\ee}{{\rm e}}

 \newcommand{\ket}[1]{|#1\rangle}
 \newcommand{\bra}[1]{\langle #1|}

\begin{document}

\title{Topological phase transitions  in the non-Abelian honeycomb lattice}

\author{A. Bermudez 
}
\affiliation{
Departamento de F\'isica Te\'orica I,
Universidad Complutense, 
28040 Madrid, 
Spain
}

\author{
N. Goldman
}
\affiliation{Center for Nonlinear Phenomena and Complex Systems - Universit$\acute{e}$ Libre de Bruxelles (U.L.B.), Code Postal 231, Campus Plaine, B-1050 Brussels, Belgium}

\author{A. Kubasiak 
}
\affiliation{
ICFO-Institut de Ci\`encies Fot\`oniques,
Parc Mediterrani de la Tecnologia,
E-08860 Castelldefels (Barcelona), Spain}

\affiliation{Marian Smoluchowski Institute of Physics Jagiellonian University, Reymonta 4, 30059 Krak\'ow, Polska
}

\author{ 
M. Lewenstein}

\affiliation{
ICFO-Institut de Ci\`encies Fot\`oniques,
Parc Mediterrani de la Tecnologia,
E-08860 Castelldefels (Barcelona), Spain}
\affiliation{
ICREA - Instituci\' io Catalana de Recerca i Estudis Avan{\c c}ats, 08010 
Barcelona, Spain}
\author{M.A. Martin-Delgado}

\affiliation{
Departamento de F\'isica Te\'orica I,
Universidad Complutense, 
28040 Madrid, 
Spain
}


\begin{abstract}
Ultracold Fermi gases trapped in honeycomb optical lattices provide an intriguing scenario where relativistic quantum electrodynamics can be tested. Here, we generalize this system to non-Abelian quantum electrodynamics, where massless Dirac fermions interact with effective non-Abelian gauge fields. We show how in this setup a variety of topological phase transitions occur, which arise due to massless fermion pair production events, as well as pair annihilation events of two kinds:
spontaneous and
strongly-interacting induced. Moreover, such phase transitions  can be controlled and characterized in optical lattice experiments.
\end{abstract}





\maketitle

\section{Introduction}

The ability to control systems at the quantum level offers an alternative and exciting route to deepen our understanding of nature~\cite{q_simulation}. In particular, the experimental design of many-body Hamiltonians provides a controlled method to study phenomena which were only believed to occur in the realm of condensed-matter physics.   Ultracold atoms  in optical lattices constitute a rich playground where the behavior of condensed-matter systems can be mimicked~\cite{qs_cond_mat_OL1,qs_cond_mat_OL2}, such as the superfluid-Mott insulator transition  for bosons~\cite{mott_greiner} and fermions~\cite{mott_fermion_1,mott_fermion_2}. Often, the experimental versatility  surpasses the possibilities of real materials, envisaging scenarios that were previously thought of as mere fiction.  For instance,  neutral bosonic or fermionic matter can be subjected to effective non-Abelian gauge fields~\cite{lewenstein_non_Abelian,spielman}. In this work, we shall describe the properties of the non-Abelian honeycomb lattice, namely, a 2D free Fermi gas in a honeycomb lattice subjected to additional non-Abelian gauge fields. The bare Fermi gas can be considered to be a quantum-optical analogue of graphene~\cite{duan_graphene_ol}, an interesting material with relativistic massless excitations that has been recently synthesized in a laboratory~\cite{graphene_exp}. As shown below, the addition of external non-Abelian gauge fields induces a variety of topological phase transitions  beyond the usual  Landau symmetry-breaking paradigm~\cite{wen_book}, which can be observed in  cold-atom experiments.    

The very unusual properties of graphene, a single layer of carbon atoms packed in a honeycomb lattice,  rely on the fact that the low-energy excitations display a linear dispersion relation~\cite{graphite}, and are thus described by massless Dirac fermions~\cite{semenoff}.  Accordingly, it is possible to observe exotic effects in low-temperature table-top experiments, which usually belong to high-energy physics (see~\cite{graphene_review} and references therein). In this context, phenomena such as Klein tunneling~\cite{grap_klein_theo,grap_klein_exp}, or the relativistic extension of  Landau levels~\cite{rel_landau_levels, rel_landau_levels_graph, grap_landau_levels_exp} have already been observed. Interestingly,  such effects might lead to the experimental realization of fully-relativistic Schr\"{o}dinger cat states, the so-called Dirac cat states~\cite{cat_states}. Besides,  the transport properties of graphene  are determined by the underlying relativistic excitations, which are responsible for the anomalous half-integer quantum Hall effect (QHE)~\cite{jackiw_qhe,schakel_qhe,sharapov_qhe,guinea_qhe,grap_anomalous_qhe_exp1,grap_anomalous_qhe_exp2}. These effects were also discussed in the context of ultracold atoms in honeycomb~\cite{duan_graphene_ol}, and $T_3$ (rhombic) lattices~\cite{bercioux_rombic}.

In this work, we describe the novel effects that take place in fermionic optical lattices when external gauge fields are switched on. Let us note that additional Abelian gauge potentials \cite{spielman,OL_Abelian_field,OL_Abelian_field_2,OL_Abelian_field_3,OL_ab_field_dark_states_1,OL_ab_field_dark_states_2, polini,polini2,Holland1,Holland2,Tung,imme} lead to analogues of well-known effects, such as the Hofstadter butterfly~\cite{OL_Abelian_field}, the Escher staircase~\cite{OL_Abelian_field_2}, the integer QHE~\cite{OL_iqhe,nathan-book}, the anomalous QHE in the honeycomb~\cite{anomalous_qhe_ol} and Kondo lattices~\cite{anomalous_qhe_kondo}, or even the fractional QHE~\cite{OL_fqhe}. When the external field is non-Abelian \cite{lewenstein_non_Abelian,fleischhauer_non_ab,cavity_non_Abelian}, a whole plethora of new phenomena arises. In this regime,  it is possible to realize a quantum-optical spin Hall effect~\cite{non_ab_spin_hall_effect}, to observe a modified Metal-Insulator transition~\cite{non_ab_MI_transition}, or to induce quasi-relativistic effects~\cite{non_ab_quasi_relat}, such as {\it Zitterbewegung}~\cite{non_ab_zitterbewegung}. It is also possible to observe non-Abelian effects in atom optics~\cite{non_ab_neg_refraction}, to obtain an optical-lattice version of a spin field-effect transistor~\cite{non_ab_spin_transisitor},  to ascertain the absence of localization in disordered relativistic one-dimensional systems~\cite{non_ab_non_localization}, or to study the effects of non-Abelian magnetic monopoles~\cite{non_ab_monopole} and Yang Mills theories~\cite{pachos}. Such proposals~\cite{lewenstein_non_Abelian,fleischhauer_non_ab} can also be  used to produce effective spin-orbit interactions~\cite{so_coupling},  to realize topological order~\cite{top_order_p_ip_sf,top_order_p_ip_sf_2}, or to generalize the integer QHE to a non-Abelian scenario~\cite{OL_non_ab_qhe}. Let us finally remark that non-Abelian features also provide an interesting setup where  Dirac fermions emerge in a square lattice~\cite{dirac_fermions_non_ab_square,nathan-book}, a regime previously restricted to staggered  fields~\cite{dirac_fermions_square_staggered_1,dirac_fermions_square_staggered_2}.

It is the purpose of the present article to study the effects of non-Abelian gauge fields on the emerging relativistic fermions that arise in the optical-lattice analogue of graphene.  As discussed below,  these fields lead to spontaneous and stimulated processes, where pairs of relativistic fermion-antifermion are created or annihilated.  Intriguingly, these interacting scattering processes occur in a non-interacting Fermi gas, which clearly shows that the background non-Abelian field has dramatic consequences on the emerging low-energy theory. Interestingly, these creation-annihilation processes can be understood in the light of topological quantum phase transitions~\cite{wen_book}, since the number of conical singularities and thus  the topology of the Fermi surface is altered. Additionally, the creation-annihilation events occur between Dirac fermions with opposite topological charges~\cite{wen_zee}, which allow us to further characterize the different topological phases.  An emphasis is placed upon the detection of this effect from density measurements.

This paper is organized as follows, in Sec.~\ref{Abelian_honeycomb} we review the properties of a 2D Fermi gas in a honeycomb optical lattice. In Sec.~\ref{non_Abelian_honeycomb}, we introduce an additional non-Abelian gauge field, and describe its effect on the emerging relativistic fermions.  Here, new phases with different Fermi surface topologies arise, and the gauge-induced phase transitions between them are presented in Sec.~\ref{tqpt}. In Sec.~\ref{QHEsect}, we present a detailed study of the anomalous quantum Hall effect that arises in the different topological phases. In Sec.~\ref{detection}, we discuss different techniques to experimentally detect and characterize these phases. Finally, Sec.~\ref{conclusions} contains our conclusions.

\section{ Abelian honeycomb lattice}
\label{Abelian_honeycomb}

\begin{figure}[!hbp]
\centering
\begin{overpic}[width=8.0cm]{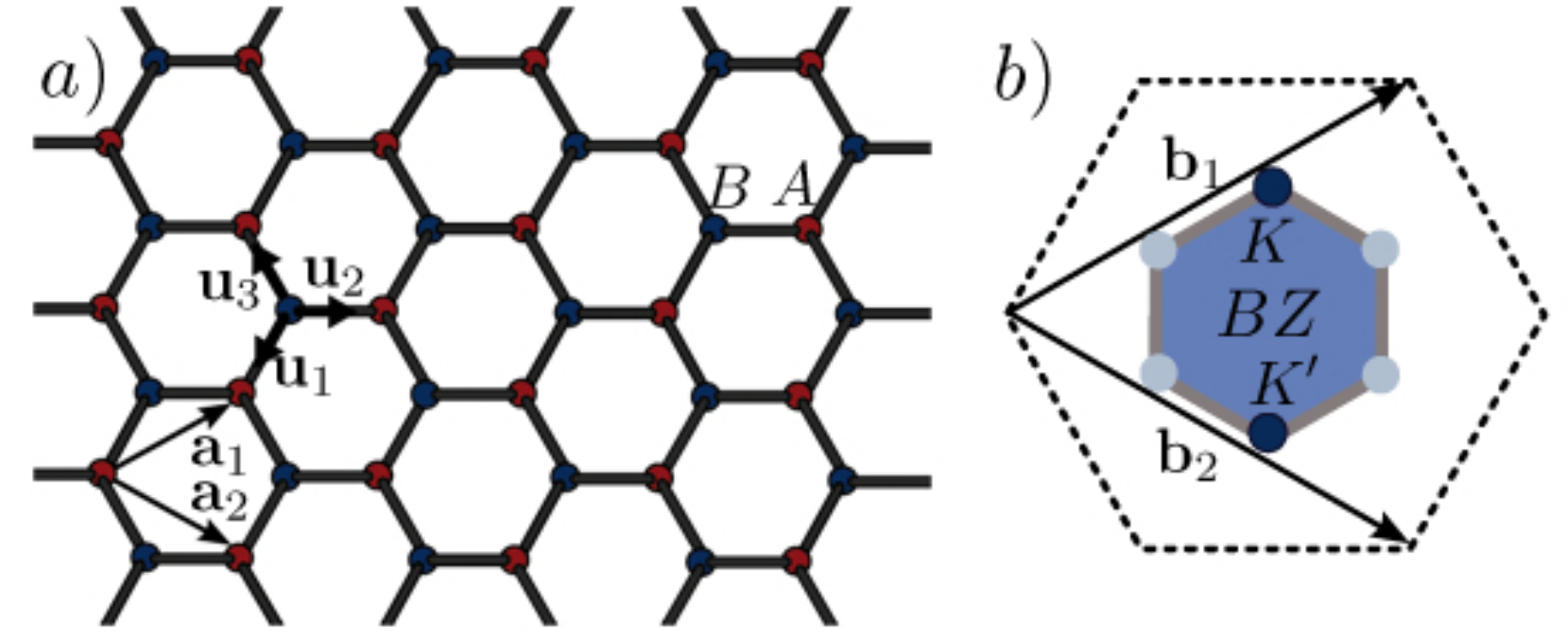}
\end{overpic}
\caption{(a) The honeycomb lattice can be described as two triangular Bravais lattices A, and B, spanned by the unit vectors $\textbf{a}_1$ and $\textbf{a}_2$. The nearest neighbors of the  B-sublattice are A-sublattice sites  given by $\textbf{u}_1,\textbf{u}_2,\textbf{u}_3$. (b) The reciprocal lattice is spanned by $\textbf{b}_1$, $\textbf{b}_2$, and the first Brillouin zone corresponds to a rotated hexagon, whose corners correspond to Dirac points (Note that only $K,K'$ are independent). }\label{fig1}
\end{figure}

In this section, we review the peculiar properties of a free Fermi gas loaded in a honeycomb lattice (see Fig.~\ref{fig1}(a)). In condensed matter, graphene provides an ideal realization of this system, where conduction electrons are tightly bound to the carbon atoms  distributed along a honeycomb crystal lattice~\cite{graphite,graphene_review}. The recent isolation of graphene~\cite{graphene_exp} reveals an experimental path towards truly 2D-materials which present distinctive features from their 3D-counterparts. In particular, graphene has the band structure of a semi-metal, since the valence and conduction bands touch at isolated points of the Brillouin zone. Moreover, the dispersion relation around such points is linear,  and   thus  low-energy excitations behave as relativistic massless Dirac fermions. In accordance, this noteworthy material connects two different fields:  relativistic quantum electrodynamics and condensed-matter physics.

A quantum-optical analogue of this 2D material can be achieved with fermionic atoms, such as $^{40}\text{K}$ or $ ^{6}\text{Li}$~\cite{mott_fermion_1,mott_fermion_2}, loaded in a honeycomb optical lattice~\cite{duan_graphene_ol} (see Fig.~\ref{fig1}(a)). The corresponding optical dipole potential  can be generated by three coplanar lasers in standing-~\cite{duan_graphene_ol} or traveling-wave configurations~\cite{dirac_fermions_OL}. In fact, the bosonic superfluid-Mott insulator transition in both honeycomb and triangular 2D lattices has been recently observed~\cite{sengstock}. In order to achieve the half-filling and non-interacting regime, pairwise interactions should be controlled by means of Feschbach resonances~\cite{feschbach}. Let us remark that this setup offers an optical version of free-standing graphene, where dislocations, impurities, phonons, or curved defects are completely absent. Furthermore, the controllability of the hopping strengths and on-site energies, leads to anisotropic and massive regimes that go beyond the real material possibilities~\cite{duan_graphene_ol,non_ab_spin_hall_effect}. 

 \subsection{Band structure: Tight-binding model and conical singularities}
\begin{figure}[!hbp]
\centering
\begin{overpic}[width=5.0cm]{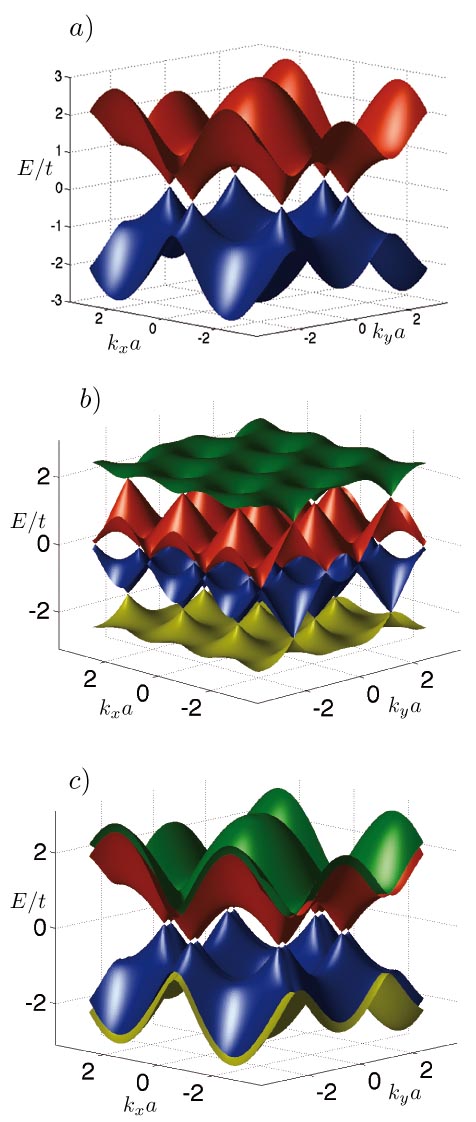}
\end{overpic}
\caption{Band structure and conical singularities of the Fermi gas in a honeycomb lattice. a) Graphene regime, where the bands  touch at the corners of the hexagonal Brillouin zone. Accordingly, the fermion gas can be described as a semiconductor with vanishing gap or a semi-metal. b)  $\Pi$-flux regime $\alpha=\beta=\pi/2$,  where the number of singularities  is clearly raised with respect to graphene. c)  Non-Abelian regime $\alpha=\beta=0.2$, where the valence and conduction bands split in two, and the structure of the conical singularities is modified as a consequence of  the  coupling to the gauge field.}
\label{fig2}

\end{figure}

\begin{figure}[!hbp]
\centering
\begin{overpic}[width=8.0cm]{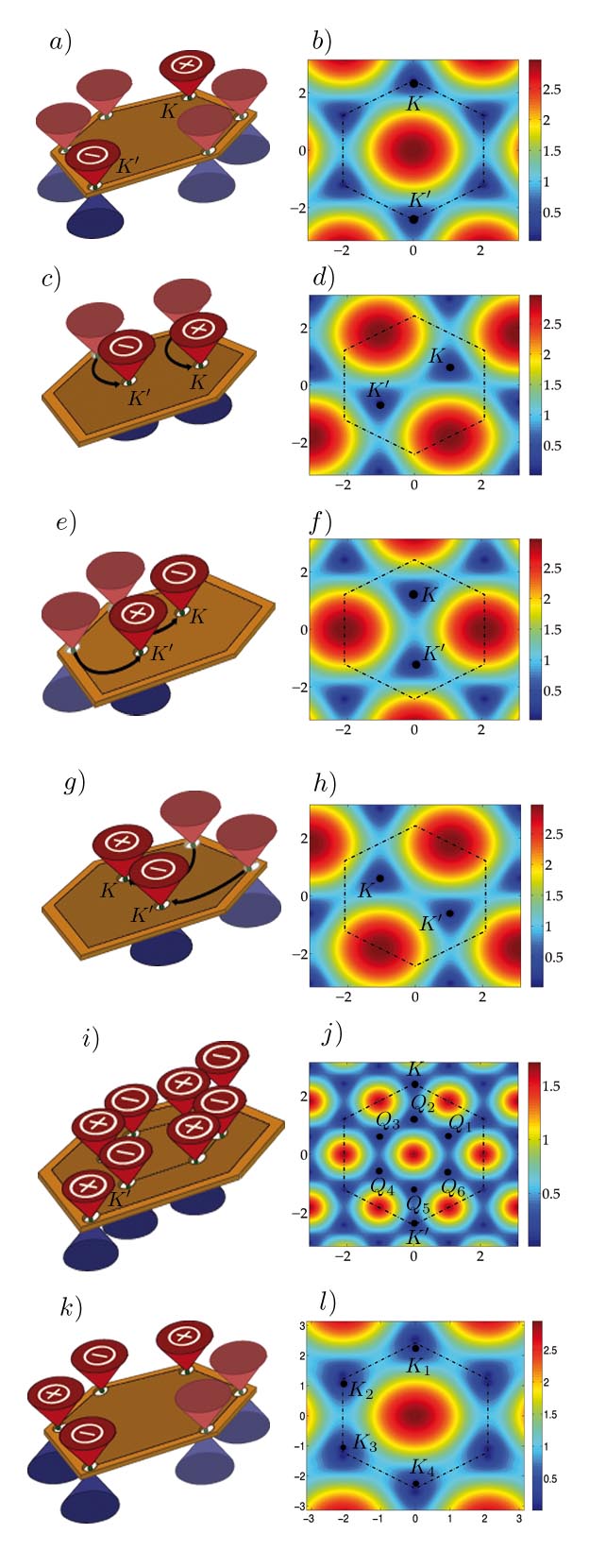}
\end{overpic}
\caption{ Distribution of Dirac points with their associated topological charges along the first Brillouin zone for (a) Graphene $P_1$, (c) ${R}_{\textbf{u}_3}$-reversed graphene $P_2$, (e) ${R}_{\textbf{u}_3}{R}_{\textbf{u}_1}$-reversed graphene  $P_3$, (g) ${R}_{\textbf{u}_1}$-reversed graphene $P_4$, (i) $\Pi$-flux $P_5$, (k) Non-Abelian point $\alpha=\beta\ll 1$. Contour plot of the energy bands for (b) Graphene $P_1$, (d) ${R}_{\textbf{u}_3}$-reversed graphene $P_2$, (f) ${R}_{\textbf{u}_3}{R}_{\textbf{u}_1}$-reversed graphene  $P_3$, (h) ${R}_{\textbf{u}_1}$-reversed graphene $P_4$, (i) $\Pi$-flux $P_5$, (l) Non-Abelian point $\alpha=\beta\ll 1$.}
\label{fig3}
\end{figure}

 Let us review the band structure of fermions hopping along the honeycomb lattice~\cite{graphite} (cf. Fig.~\ref{fig1}(a)), which has a two-atom basis with unit vectors  $\textbf{a}_1=\textstyle{\frac{a}{2}}(3,\sqrt{3})$ and $\textbf{a}_2=\textstyle{\frac{a}{2}}(3,-\sqrt{3})$, and lattice spacing $a$. Fermions in the A-sublattice hop to three nearest neighbors of the B-sublattice $\textbf{u}_1=-\textstyle{\frac{a}{2}}(1,\sqrt{3})$, $\textbf{u}_2=a(1,0)$, and $\textbf{u}_3=\textstyle{\frac{a}{2}}(-1,\sqrt{3})$, according to the following Hamiltonian
 \begin{equation}
 \label{Abelian_hopping}
 H=-t\sum_{\langle i,j\rangle}a^{\dagger}_ib_j+\text{h.c.}
 \end{equation}
Here, the fermionic operators $a^{\dagger}_i,b^{\dagger}_i$ ($a_i,b_i$) create (annihilate) a fermion on $i-$th site of the A- and B-sublattice respectively, and $t$ stands for the nearest-neighbor hopping energy. Note that the hopping strength is controlled through the optical potential depth~\cite{dirac_fermions_OL}. The translationally invariant Hamiltonian in Eq.~\eqref{Abelian_hopping} can be diagonalized in momentum space by the introduction of  sublattice operators $a_{\textbf{k}}=\sum_{\textbf{r}_j\in A}\ee^{\ii\textbf{k}\textbf{r}_j}a_j $, and $ b_{\textbf{k}}=\sum_{\textbf{r}_j\in B}\ee^{\ii\textbf{k}\textbf{r}_j}b_j$. Accordingly, the Hamiltonian becomes a quadratic operator
$H=\sum_{\textbf{k}}\Psi_{\textbf{k}}^{\dagger}H_{\textbf{k}}\Psi_{\textbf{k}}$, where
\begin{equation}
\label{momentum_tb_hamiltonian}
H_{\textbf{k}}=-t\sum_{j}\left(\ee^{\ii\textbf{k}\textbf{u}_j}\sigma^++\ee^{-\ii\textbf{k}\textbf{u}_j}\sigma^-\right)
\end{equation}
and  $\Psi_{\textbf{k}}=(a_{\textbf{k}},b_{\textbf{k}})^{t}$ is a spinor operator. Note that the Hamiltonian in Eq.~\eqref{momentum_tb_hamiltonian} is expressed in terms of the sublattice flip operators $\sigma^+=\ket{\!\uparrow}\bra{\downarrow\!}$, $\sigma^-=\ket{\!\downarrow}\bra{\uparrow\!}$, with $\uparrow$ ($\downarrow$) representing the A(B)-sublattice. This Hamiltonian is readily diagonalized, yielding the following band structure
\begin{equation}
\label{Abelian_energy_bands}
E_{\pm}(\textbf{k})=\pm t\sqrt{3+4\cos\textstyle{\frac{3k_xa}{2}}\cos\textstyle{\frac{\sqrt{3}k_ya}{2}}+2\cos\sqrt{3}k_ya},
\end{equation}
which is represented in Fig.~\ref{fig2}(a). In this figure, we observe that the two energy bands touch at different points inside the first Brillouin zone, and therefore the Fermi surface at half filling is reduced to a discrete set of isolated points (see also Fig.~\ref{fig1}(b)). Besides, a closer inspection shows that the energy dispersion around these singular points is conical $E\sim|\textbf{k}|$ (see Fig.~\ref{fig3}(a)), which is a clear signature of the relativistic nature of the low-energy excitations at half filling. 

Let us now describe the content of figs.~\ref{fig3}(a) and~\ref{fig3}(b) in more detail. In Fig.~\ref{fig3}(a), we schematically represent the distribution of Dirac cones inside the hexagonal first Brillouin zone. We find six conical singularities localized on the corners of the Brillouin zone, where the valence and conduction bands touch (see also Fig.~\ref{fig2}(a)). Note that only two of such Dirac points  are independent, the remaining being obtained from this pair by mere translations in the reciprocal unit basis  $\textbf{b}_1=\textstyle{\frac{2\pi}{3a}}(1,\sqrt{3})$, $\textbf{b}_2=\textstyle{\frac{2\pi}{3a}}(1,-\sqrt{3})$. The two independent Dirac points $\textbf{K},\textbf{K}'$ are represented by opaque cones, whereas the remaining cones are transparent. Accordingly, we can state that two conical singularities arise in the band structure of the honeycomb lattice. This schematic figure is further supported by Fig.~\ref{fig3}(b), where the contour of the conduction band in Eq.~\eqref{Abelian_energy_bands} has been represented. In this contour  plot, the color scale is the following: dark blue colors represent energies close to zero $E_{F}=0$, while light red colors correspond to higher energies, and thus to empty fermionic states. From this figure, we observe that six zero-energy points arise in the corners of the Brillouin zone, an hexagon represented with dashed lines. It is precisely at these zero-energy points that the conduction and valence bands touch, giving rise to the conical singularities formerly described. As shown in the next section,  massless Dirac fermions emerge as the fundamental low-energy excitations around these singular points.

\subsection{Low-energy excitations: relativistic  fermions}

To study the low-energy excitations,  we  expand the Hamiltonian in Eq.~\eqref{momentum_tb_hamiltonian} around the conical singularities located at  $\textbf{K}=\textstyle{\frac{2\pi}{3a}}(0,\textstyle{\frac{2}{\sqrt{3}}})$, $\textbf{K}'=\textstyle{\frac{2\pi}{3a}}(0,-\textstyle{\frac{2}{\sqrt{3}}})$. The long-wavelength excitations  are obtained after expanding  the momentum around each Dirac point, such as $\textbf{K}$, namely $\textbf{p}=\hbar(\textbf{k}-\textbf{K})$, which leads to
\begin{equation}
\label{Abelian_dirac_hamiltonian}
H_{\text{eff}}=\sum_{\textbf{p}}\Psi_{\textbf{p}}^{\dagger},H_{\text{D}}(\textbf{p})\Psi_{\textbf{p}},\hspace{0.5ex}H_{\text{D}}= c\left(\alpha_xp_x+\alpha_yp_y\right),
\end{equation}
where $\alpha_x=\sigma_y$, $\alpha_y=-\sigma_x$, are the Dirac matrices,  which   are reduced to the usual Pauli matrices in a 2+1 Minkowski spacetime. This Hamiltonian is completely equivalent to the Dirac Hamiltonian of massless fermions, if we understand the Fermi velocity $c=\textstyle{\frac{3}{2}\frac{ta}{\hbar}}$ as the effective speed of light. The relativistic Hamiltonian in Eq.~\eqref{Abelian_dirac_hamiltonian} readily accounts for the conical singularities $E\sim c|\textbf{p}|$ observed in Fig.~\ref{fig2}(a). Additionally, it identifies the lattice geometry (i.e. the existence of two interpenetrating triangular sublattices) as the responsible of the spinor structure of the effective relativistic theory. Let us note that the low energy quantum field theory associated to the remaining Dirac point $\textbf{K}'$ also describes emerging massless fermions, but with a different representation of the Dirac matrices (i.e. satisfying the Clifford algebra $\{\alpha_j,\alpha_k\}=2\delta_{jk}$), namely, $\alpha_x=\sigma_y$, $\alpha_y=\sigma_x$. Therefore, two species of independent  Dirac fermions $N_{d}=2$ emerge in the low-energy theory of a two-dimensional Fermi gas in the half-filled lattice. 

\begin{figure}[!hbp]
\centering
\begin{overpic}[width=8.50cm]{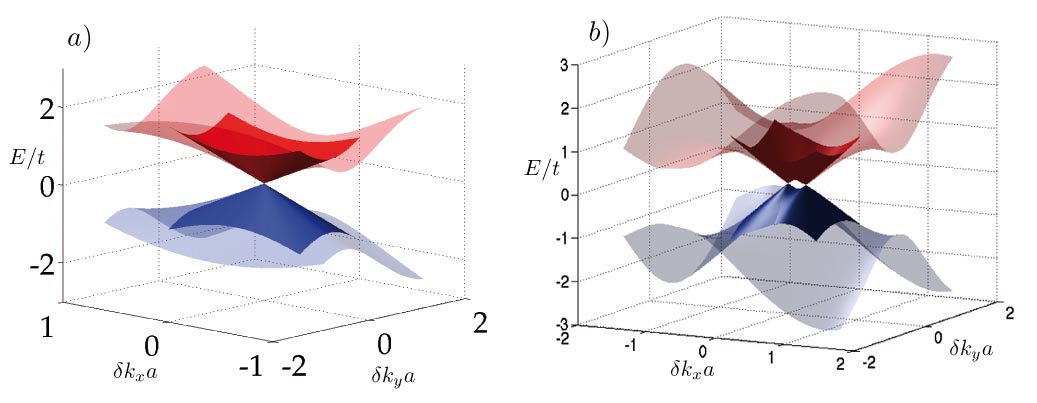}
\end{overpic}
\caption{Comparison between exact energy bands and the low-energy approximation. a) In graphene, we represent  the exact energies in Eq.~\eqref{Abelian_energy_bands} (transparent colors) and the energy spectrum from the low-energy Hamiltonian in Eq.~\eqref{Abelian_dirac_hamiltonian} (solid colors) around $\textbf{K}$. b)  Idem for $\alpha=\beta=0.2$,  comparing  Eq.~\eqref{non_Abelian_bands} to Eq.~\eqref{Abelian_dirac_hamiltonian} around $\textbf{K}_1$.}
\label{low_energy_graphene_bands}
\end{figure}

The effective relativistic picture emerging at low energies in Eq.~\eqref{Abelian_dirac_hamiltonian} should be compared with the exact energy bands in Eq.~\eqref{Abelian_energy_bands}. In Fig.~\ref{low_energy_graphene_bands}(a), we represent the exact band structure around the Dirac point \textbf{K} (transparent surfaces), and compare it with the conical energy dispersion predicted from the Dirac Hamiltonian (solid surfaces). Accordingly, we show that the description of low-energy excitations in terms of relativistic massless fermions is valid in a neighborhood of the Dirac points (i.e.$\delta\textbf{k}=\textbf{k}-\textbf{K}\ll1/a$).

\section{ Non-Abelian honeycomb lattice}
\label{non_Abelian_honeycomb} 

In the previous section, we derived the effective low-energy quantum field theory for the half-filled fermionic honeycomb lattice, and showed that two species of massless Dirac fermions emerge at large wavelengths, each around a different Dirac point. The coupling of such massless fermions to external  electromagnetic fields leads to interesting effects distinctive of  relativistic field theories, such as the Klein paradox~\cite{grap_klein_exp}. In this section, we shall focus on  the fermion coupling to constant magnetic fields. In the Abelian scenario, this leads to the noteworthy anomalous integer quantum Hall effect, either in graphene~\cite{jackiw_qhe,schakel_qhe,sharapov_qhe,guinea_qhe,grap_anomalous_qhe_exp1,grap_anomalous_qhe_exp2}, or honeycomb optical lattices~\cite{anomalous_qhe_ol}.  As discussed below, this paradigm should be reconsidered when matter is coupled to non-Abelian gauge potentials. Let us remark that even if such fields are not available in condensed-matter experiments, it is possible to synthesize them in a controlled manner by means of laser-assisted tunneling processes in optical lattices~\cite{lewenstein_non_Abelian}, or dark-state methods~\cite{fleischhauer_non_ab}. Therefore, surpassing graphene-based materials, optical lattices present exotic avenues at the forefront of condensed matter, high-energy physics, and atomic physics.

\begin{figure}[!hbp]
\centering
\begin{overpic}[width=6.50cm]{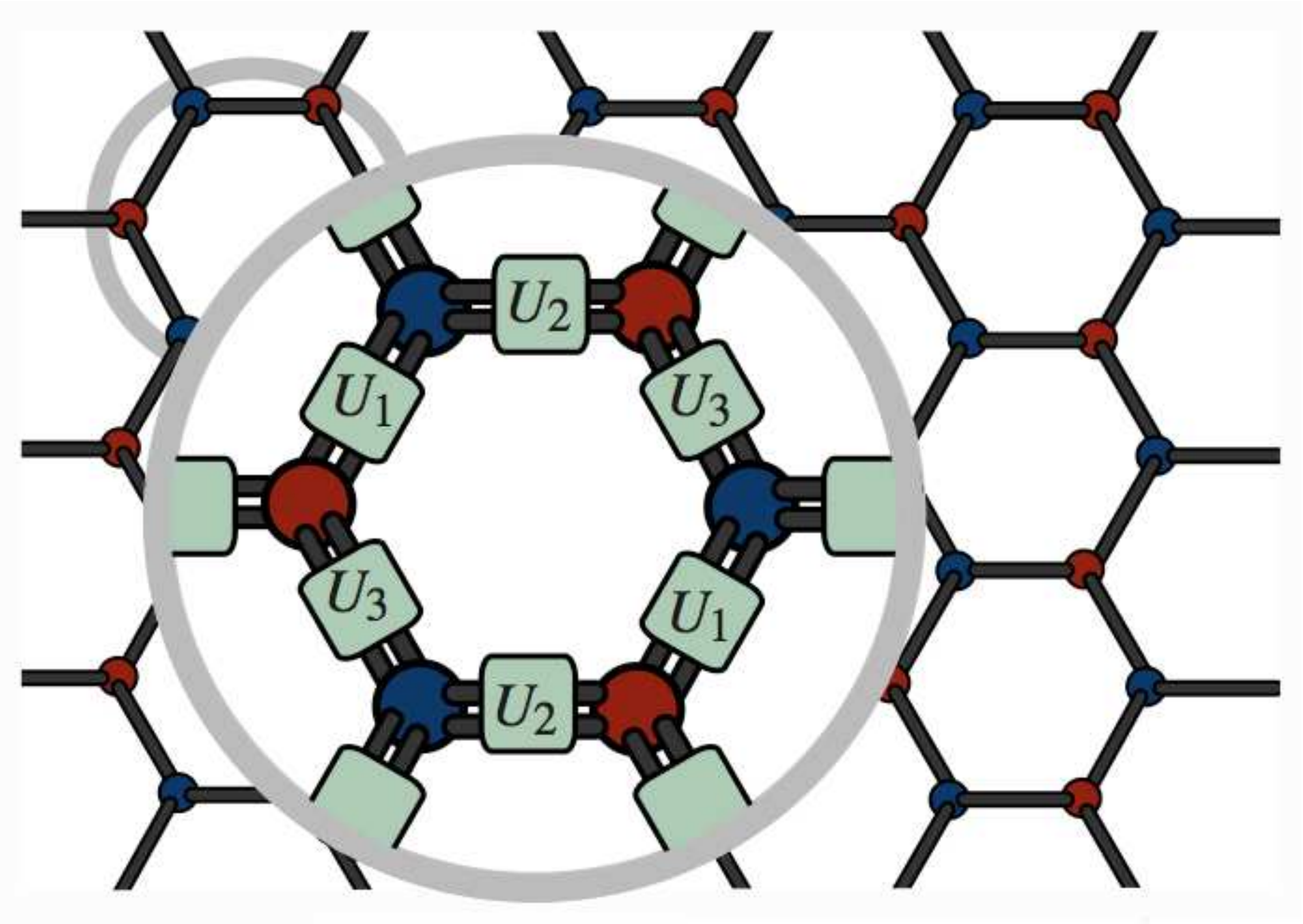}
\end{overpic}
\caption{ Scheme for the fermionic honeycomb lattice subjected to SU(2) gauge fields, where each hopping is dressed by  $U_1=\ee^{\ii\alpha\tau_x}$, $U_2=\mathbb{I}$, and $U_3=\ee^{\ii\beta\tau_y}$. We have used  double-line links to indicate  the underlying color degrees of freedom $\tau=1,2$ associated to each fermion. The gauge fields on the links induce a color-dependent hopping between adjacent  sites, which has been represented by squares. }
\label{non_ab_honeycomb}
\end{figure}

In non-Abelian lattice gauge theory,  a unitary matrix $U_{ij}$ is associated  to the link connecting the lattice points $\textbf{r}_i\rightarrow\textbf{r}_j$, which must be generated by the elements of the underlying Lie group. Consequently, the Abelian hopping Hamiltonian in Eq.~\eqref{Abelian_hopping} must be generalized to
 \begin{equation}
 \label{non_Abelian_hopping}
 H=-t\sum_{\langle i,j\rangle}\left[U_{ij}\right]_{\tau\tau'}a^{\dagger}_{j\tau'}b_{i\tau}+\text{h.c.},
 \end{equation}
where $a^{\dagger}_{i\tau}$, $b^{\dagger}_{i\tau}$ ($a_{i\tau}$, $b_{i\tau}$) create (annihilate) a fermion at lattice site  $\textbf{r}_i$ of the A- and B-sublattice, with a color-like component $\tau=1,2,...,N$ corresponding to the fundamental representation of the Lie group. Such a situation has been depicted in Fig.~\ref{non_ab_honeycomb} for the particular case of SU(2) fields (i.e. fermions have two colors $\tau=1,2$ ). Here, we observe that the coupling of matter to non-Abelian fields induces  unitary transformations along each hopping path: $U_1=\ee^{\ii\alpha\tau_x}$, $U_2=\mathbb{I}$, and $U_3=\ee^{\ii\beta \tau_y}$, where $\tau_x, \tau_y$ are Pauli matrices in the color-components, and $\alpha,\beta$ represent the non-Abelian fluxes (see Fig.~\ref{non_ab_honeycomb}). In the momentum representation, the Hamiltonian in Eq.~\eqref{non_Abelian_hopping} becomes $H=\sum_{\textbf{k}}\Psi_{\textbf{k}}^{\dagger}H_{\textbf{k}}\Psi_{\textbf{k}}$, where 
\begin{equation}
\label{non_Abelian_momentum_hamiltonian}
H_{\textbf{k}}=-t\sum_{j}\left(\ee^{\ii\textbf{k}\textbf{u}_j}\sigma^+\otimes U_{j}+\ee^{-\ii\textbf{k}\textbf{u}_j}\sigma^-\otimes U^{\dagger}_j\right),
\end{equation}
and  the spinor $\Psi_{\textbf{k}}=(a_{\textbf{k}1},a_{\textbf{k}2},b_{\textbf{k}1},b_{\textbf{k}2})^{t}$  includes pseudo-spin and color degrees of freedom, which correspond to the underlying sublattices and to the internal fermionic components. 
 
In the following sections, we study the pattern of emerging relativistic fermions associated to different gauge fluxes $(\alpha,\beta)$. As shown below, the total number of  conical singularities and their distribution within the Brillouin zone depends on the particular values of the gauge fluxes. We shall characterize different Abelian and non-Abelian phases  in terms of these low-energy massless fermions.

\subsection{Wilson loop and non-Abelian regimes}
\begin{figure}[!hbp]
\centering
\begin{overpic}[width=5.0cm]{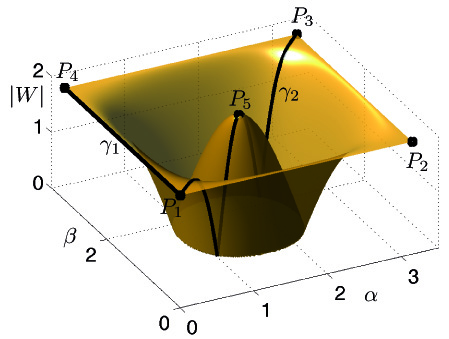}
\end{overpic}
\caption{Wilson loop associated to the  hexagonal plaquette under the gauge fields in Eq.~\eqref{gauge_fields}. The Abelian regimes ($|W|=2$), correspond to  the  flux configurations $P_1=(0,0)$, $P_2=(0,\pi)$, $P_3=(\pi,\pi)$, $P_4=(\pi,0)$, the straight lines joining them, and the marginally Abelian point $P_5=(\frac{\pi}{2},\frac{\pi}{2})$.  Different paths, such as $\gamma_1$ and $\gamma_2$, join the above regimes by means of a quench in the external fluxes. }\label{wilson_loop}

\end{figure}

The non-Abelian gauge fields that induce these conditional hoppings are defined through a generalized Peierls substitution $U_{ij}=V^{\dagger}_i\ee^{\ii a A_{\mu}(\textbf{r}_{\text{l}})}V_j$, where the independent matrices $V_j$ reflect the local gauge invariance, and the gauge field $A_{\boldsymbol{\mu}}(\textbf{r}_{\text{l}})$ is located at the link $\textbf{r}_{\text{l}}=\half(\textbf{r}_i+\textbf{r}_j)$ and directed towards $\boldsymbol{\mu}=\textbf{r}_j-\textbf{r}_i$.  The particular choice $U_1=\ee^{\ii\alpha\tau_x}$, $U_2=\mathbb{I}$, and $U_3=\ee^{\ii\beta\tau_y}$ in Fig.~\ref{non_ab_honeycomb}, is locally equivalent to the following SU(2)-generated  gauge field $\textbf{A}=\sum_{a\mu}A^{a}_{\mu}(\textbf{r}_{\text{l}})\tau_{a}\textbf{e}_{\mu}$
\begin{equation}
\label{gauge_fields}
\textbf{A}=\alpha(\tau_x-\tau_y)\textbf{e}_x+\frac{\beta}{\sqrt{3}}(\tau_x+\tau_y)\textbf{e}_{y}.
\end{equation}
In the continuum limit $a\to 0$, and according to non-Abelian gauge theory, the associated strength tensor $F_{\mu\nu}=\partial_{\mu}A_{\nu}-\partial_{\nu}A_{\mu}-\ii [A_{\mu},A_{\nu}]$, describes a  constant magnetic field  along the $z-$axis $\textbf{B}=\frac{2\alpha\beta}{\sqrt{3}}\tau_z\textbf{e}_z$. On the lattice, these external fields are further characterized by the  Wilson loop $W=\text{tr} U_{\circlearrowleft}$, where $U_{\circlearrowleft}=U_1U_2U_3U_1^{\dagger}U_2^{\dagger}U_3^{\dagger}$ is the  unitary transformation around an elementary plaquette. The Wilson loop represents a gauge-invariant non-local observable related to the non-Abelian flux through an hexagonal plaquette, and is expressed  as follows
\begin{equation}
W(\alpha,\beta)=2\cos^2\alpha+2\cos{2\beta}\sin^2\alpha.
\end{equation}
In Fig.~\ref{wilson_loop}, we depict the Wilson loop as a function of the external fluxes, and identify the Abelian ($|W|=2$) and non-Abelian ($|W|\neq2$) regimes.

\vspace{1ex}
 \textbf{Abelian regimes:} The following values of the gauge fluxes correspond to Abelian phases in the Hamiltonian of Eq.~\eqref{non_Abelian_hopping} 
 \begin{itemize}
\item {\bf Graphene point} $P_1$ for $(\alpha,\beta)=(0,0),$
\item $\textbf{R}_{\textbf{u}_3}${\bf-reversed graphene} $P_2$ for $(\alpha,\beta)=(0,\pi),$
 \item  $\textbf{R}_{\textbf{u}_3}\textbf{R}_{\textbf{u}_1}${\bf-reversed graphene} $P_3$ for $(\alpha,\beta)=(\pi,\pi)$,
\item $\textbf{R}_{\textbf{u}_1}${\bf-reversed graphene} $P_4$ for $(\alpha,\beta)=(\pi,0)$.
  \item {\bf $\Pi$-flux point} $P_5$ for $(\alpha,\beta)=(\half\pi,\half\pi)$.  
\end{itemize}
  Note also that the lines connecting $P_1,P_2,P_3,P_4$ correspond to Abelian phases (see Fig.~\ref{wilson_loop}). In the next section, we derive the band structure and low-energy massless fermions for these different regimes. We show that the four graphene-like points $P_1,P_2,P_3,P_4$ have $N_d=4$ species of emerging Dirac fermions (notice the doubling due to color-degeneracy), and are thus topologically equivalent. Conversely, the $\Pi$-flux phase presents $N_{d}=8$ different gapless fermions, and thus a distinct Fermi surface topology. As shown in Sec.~\ref{QHEsect}, transport properties in this $\Pi$-flux phase, such as Hall conductances,  clearly deviate from the graphene-like phases.
 
  \vspace{1ex}
 \textbf{Non-Abelian regimes:} The remaining values of the gauge fluxes define non-Abelian phases with $|W|<2$, whose low-energy properties shall dramatically depend on the particular values of the fluxes. Below, we study the non-Abelian regime $\alpha=\beta\ll1$  close to the Graphene point $P_1$, where the low energy dispersion is linear and saddle-point in orthogonal directions. In this respect, the effective mass of quasiparticles at large wavelengths is  highly anisotropic, and leads to novel features in the transport properties.

\subsection{Effective low energy theories in Abelian regimes}

In this section, we study the emerging quantum field theories of the non-Abelian honeycomb lattice at half-filling, centering our attention to the Abelian regimes introduced above.

\vspace{1ex}
{\bf  Graphene point:} In the limit of vanishing gauge fluxes $\alpha=\beta=0$ (see point $P_1$ in Fig.~\ref{wilson_loop}), we recover the standard graphene Hamiltonian of Eq.~\eqref{momentum_tb_hamiltonian}, with an additional 2-fold color degeneracy
\begin{equation}
\label{graphene_hop}
H_{\textbf{k}}=-t\left(\ee^{\ii\textbf{k}\textbf{u}_1}+\ee^{\ii\textbf{k}\textbf{u}_2}+\ee^{\ii\textbf{k}\textbf{u}_3}\right)\sigma^+\otimes\mathbb{I}_2+\text{h.c.}
\end{equation}
Therefore, energy bands correspond to those expressed in Eq.~\eqref{Abelian_energy_bands} and shown in Fig.~\ref{fig2}, with the peculiarity that each band is doubly-degenerate. As a consequence, the number of independent Dirac points is raised to $N_{{d}}=4$, fulfilling
\begin{equation}
\label{graphene_point}
H_{\text{D}}^{\textbf{K}}= c\left(\sigma_y\otimes\mathbb{I}_2p_x-\sigma_x\otimes\mathbb{I}_2p_y\right),
\end{equation}
around the singularity $\textbf{K}$, and 
\begin{equation}
H_{\text{D}}^{\textbf{K}'}= c\left(\sigma_y\otimes\mathbb{I}_2p_x+\sigma_x\otimes\mathbb{I}_2p_y\right),
\end{equation}
around $\textbf{K}'$. As expected, the species of emerging massless fermions is doubled due to the color degrees of freedom. 

\vspace{1ex}
{\bf $\textbf{R}_{\textbf{u}_3}$-reversed graphene:} In the Abelian phase corresponding to $\alpha=0,\beta=\pi$ (point $P_2$ in Fig.~\ref{wilson_loop}), the Hamiltonian 
\begin{equation}
H_{\textbf{k}}=-t\left(\ee^{\ii\textbf{k}\textbf{u}_1}+\ee^{\ii\textbf{k}\textbf{u}_2}-\ee^{\ii\textbf{k}\textbf{u}_3}\right)\sigma^+\otimes\mathbb{I}_2+\text{h.c.},
\end{equation}
can be obtained from  graphene's analogue in Eq.~\eqref{graphene_hop}, after reversing the hopping amplitude along $\textbf{u}_3$ (i.e. $t\rightarrow -t$). The energy bands fulfill
\begin{equation}
\label{120_energy_bands}
E_{\pm}(\textbf{k})=\pm t\sqrt{3-4\sin\textstyle{\frac{3k_xa}{2}}\sin\textstyle{\frac{\sqrt{3}k_ya}{2}}-2\cos\sqrt{3}k_ya},
\end{equation}
whose contour clearly shows that the pair of conical singularities are no longer at the corners of the Brillouin zone, but well inside it (see the contour energy in Fig.\ref{fig3}(d)).  The pair of conical singularities, now located at $\textbf{K}=\textstyle{\frac{2\pi}{3a}}(\textstyle{\frac{1}{2}},\textstyle{\frac{1}{2\sqrt{3}}})$, and $\textbf{K}'=\textstyle{\frac{2\pi}{3a}}(\textstyle{\frac{-1}{2}},\textstyle{\frac{-1}{2\sqrt{3}}})$, correspond to the corner Dirac points of graphene, which have been translated in momentum space along $120^{\text{o}}$ direction. Such interesting transport of Dirac cones around the Brillouin zone is schematically represented in Fig.~\ref{fig3}(c). Here, transparent cones represent the initial Dirac fermions for vanishing fluxes  $\alpha=\beta=0$, and opaque cones represent the transported cones at  $\alpha=0,\beta=\pi$. The direction of transport is shown by black arrows. The effective low energy theory at half-filling around these points is also captured by the relativistic Dirac equation, with color-degenerate massless fermions (i.e. a total of $N_{{d}}=4$ fermions), that behave according to the Hamiltonian
\begin{equation}
H_{\text{D}}^{\textbf{K}}= c\left(\alpha_x\otimes\mathbb{I}_2p_x+\alpha_y\otimes\mathbb{I}_2p_y\right),
\end{equation}
around the singularity $\textbf{K}$, where a different representation of the Clifford algebra is chosen $\alpha_x=-\tilde{\sigma}_y,\alpha_y=\tilde{\sigma}_x$, where
\begin{equation}
\label{tilde_pauli}
\tilde{\sigma}_x=\left(\begin{array}{cc}0 & w \\ w^* & 0\end{array}\right),\hspace{2ex}\tilde{\sigma}_y=\left(\begin{array}{cc}0 & -\ii w \\\ii w^* & 0\end{array}\right),
\end{equation}
and $w=\half(1-\ii\sqrt{3})$ lies in the complex unit circle.  Around $\textbf{K}'$, the low-energy Hamiltonian leads  to $\alpha_x=\tilde{\sigma_y},\alpha_y=\tilde{\sigma}_x$.

\vspace{1ex}
{\bf $\textbf{R}_{\textbf{u}_3}\textbf{R}_{\textbf{u}_1}$-reversed graphene:} Considering the gauge fluxes $\alpha=\beta=\pi$ corresponding to the point $P_3$ in Fig.~\ref{wilson_loop}, the Hamiltonian
\begin{equation}
H_{\textbf{k}}=-t\left(-\ee^{\ii\textbf{k}\textbf{u}_1}+\ee^{\ii\textbf{k}\textbf{u}_2}-\ee^{\ii\textbf{k}\textbf{u}_3}\right)\sigma^+\otimes\mathbb{I}_2+\text{h.c.},
\end{equation}
is analogous to graphene with the reversed hoppings $t\rightarrow -t$ along the directions $\textbf{u}_1,\textbf{u}_3$. The underlying energy bands are
\begin{equation}
\label{180_energy_bands}
E_{\pm}(\textbf{k})=\pm t\sqrt{3-4\cos\textstyle{\frac{3k_xa}{2}}\cos\textstyle{\frac{\sqrt{3}k_ya}{2}}+2\cos\sqrt{3}k_ya},
\end{equation}
and lead to the conical singularities depicted in figs.~\ref{fig3}(e) and \ref{fig3}(f). The pair of conical singularities lie at $\textbf{K}=\textstyle{\frac{2\pi}{3a}}(0,\textstyle{\frac{1}{\sqrt{3}}})$, and $\textbf{K}'=\textstyle{\frac{2\pi}{3a}}(0,\textstyle{\frac{-1}{\sqrt{3}}})$, and correspond to the corner singularities  translated in momentum space along a $180^{\text{o}}$ direction. At half-filling, the low energy excitations around such points are also relativistic massless fermions
\begin{equation}
H_{\text{D}}^{\textbf{K}}= c\left(\alpha_x\otimes\mathbb{I}_2p_x+\alpha_y\otimes\mathbb{I}_2p_y\right),
\end{equation}
with $\alpha_x=\sigma_y, \alpha_y=\sigma_x$ for $\textbf{K}$, and $\alpha_x=\sigma_y, \alpha_y=-\sigma_x$ for $\textbf{K}'$.

\vspace{1ex}
{\bf $\textbf{R}_{\textbf{u}_1}$-reversed graphene:} Considering the gauge fluxes $\alpha=\pi,\beta=0$ corresponding to $P_4$ in Fig.~\ref{wilson_loop}, the  Hamiltonian
\begin{equation}
H_{\textbf{k}}=-t\left(-\ee^{\ii\textbf{k}\textbf{u}_1}+\ee^{\ii\textbf{k}\textbf{u}_2}+\ee^{\ii\textbf{k}\textbf{u}_3}\right)\sigma^+\otimes\mathbb{I}_2+\text{h.c.},
\end{equation}
is analogous to graphene with the reversed hoppings $t\rightarrow -t$ along the direction $\textbf{u}_1$. The underlying energy bands become
\begin{equation}
\label{60_energy_bands}
E_{\pm}(\textbf{k})=\pm t\sqrt{3+4\sin\textstyle{\frac{3k_xa}{2}}\sin\textstyle{\frac{\sqrt{3}k_ya}{2}}-2\cos\sqrt{3}k_ya},
\end{equation}
represented in figs.~\ref{fig3}(g) and \ref{fig3}(h). The pair of singularities lie at $\textbf{K}=\textstyle{\frac{2\pi}{3a}}(\textstyle{\frac{-1}{2}},\textstyle{\frac{3}{2\sqrt{3}}})$, and $\textbf{K}'=\textstyle{\frac{2\pi}{3a}}(\textstyle{\frac{1}{2}},\textstyle{\frac{-3}{2\sqrt{3}}})$, and correspond to the corner singularities  translated in momentum space along a $60^{\text{o}}$ direction. At half-filling, the low energy excitations around such points are also relativistic massless fermions
\begin{equation}
\label{graphene_point_bis}
H_{\text{D}}^{\textbf{K}}= c\left(\alpha_x\otimes\mathbb{I}_2p_x+\alpha_y\otimes\mathbb{I}_2p_y\right),
\end{equation}
with $\alpha_x=-\tilde{\sigma}_y$, and $\alpha_y=\tilde{\sigma}_x$ for $\textbf{K}$, where $\tilde{\sigma}_j$ correspond to Eq.~\eqref{tilde_pauli} for $w=\half(1+\ii\sqrt{3})$. Equivalently, the $\textbf{K}'$ emerging fermion leads to the Clifford algebra  $\alpha_x=\tilde{\sigma}_y$, and $\alpha_y=\tilde{\sigma}_x$.

Let us finally remark the fact that all the Abelian phases described so far simply correspond to finite translations of the graphene points, and thus present the same number of emerging fermions $N_{d}=4$. Accordingly, the physical properties of the two-dimensional Fermi gas at half filling are expected to be analogous to graphene. The unique effect of a non-vanishing Abelian flux $\alpha,\beta$ is to transport the relativistic fermions around the Brillouin zone. Below, we study richer situations where the topology of the Fermi surface is modified by a variation of the number of gapless fermionic species. It is also important to note that these Abelian phases can also be achieved by simply tuning the anisotropy of the hopping parameters~\cite{duan_graphene_ol}, without the need to insert external fields. Nonetheless, the phases to be discussed below cannot be achieved in this fashion, and constitute a non-trivial consequence of the applied gauge fields. 

\vspace{1ex}
{\bf $\Pi$-flux point:} Let us now focus on the Abelian regime $\alpha=\beta=\textstyle{\frac{\pi}{2}}$, which corresponds to the $\Pi$-flux phase (i.e. lattice fermions pick up an overall minus sign by hopping around the elementary plaquette). This regime corresponds to the point $P_5$ in Fig.~\ref{wilson_loop}, where the fermionic  Hamiltonian 
\begin{equation}
H_{\textbf{k}}=-t\left(\ii\ee^{\ii\textbf{k}\textbf{u}_1}\sigma^+\otimes\tau_x+\ee^{\ii\textbf{k}\textbf{u}_2}\sigma^+\otimes\mathbb{I}_2+\ii\ee^{\ii\textbf{k}\textbf{u}_3}\sigma^+\otimes\tau_y\right)+\text{h.c.},
\end{equation}
induces a color flip transformation when the fermion hops along  directions $\textbf{u}_1$ and $\textbf{u}_3$. The energy bands
\begin{equation}
\label{pi_flux_energy_bands}
E(\textbf{k})=\pm t\sqrt{3\pm\sqrt{6-4\cos\textstyle{3k_xa}\cos\textstyle{\sqrt{3}k_y}-2\cos2\sqrt{3}k_ya}},
\end{equation}
have been represented in Fig.~\ref{fig2}(b), where one observes how the number of conical singularities is increased with respect to the previous Abelian phases. Let us  note that the  color-degeneracy is  lifted due to the color-flip induced hopping. The distribution of conical singularities in the Brillouin zone can be better appreciated in figs.~\ref{fig3}(i) and \ref{fig3}(j), where $N_{{d}}=8$ independent Dirac points appear in  positions
\begin{equation}
\begin{split}
&\textbf{K}=\frac{2\pi}{3a}\left(0,\frac{2}{\sqrt{3}}\right),\hspace{2.5ex}\textbf{K}'=\frac{2\pi}{3a}\left(0,\frac{-2}{\sqrt{3}}\right),\\
&\textbf{Q}_1=\frac{\pi}{3a}\left(1,\frac{1}{\sqrt{3}}\right),\hspace{1.7ex}\textbf{Q}_2=\frac{\pi}{3a}\left(0,\frac{2}{\sqrt{3}}\right), \\
&\textbf{Q}_3=\frac{\pi}{3a}\left(-1,\frac{1}{\sqrt{3}}\right),
\hspace{0.3ex}\textbf{Q}_4=\frac{\pi}{3a}\left(-1,\frac{-1}{\sqrt{3}}\right),\\
&\textbf{Q}_5=\frac{\pi}{3a}\left(0,\frac{-2}{\sqrt{3}}\right),\hspace{2.2ex}\textbf{Q}_6=\frac{\pi}{3a}\left(1,\frac{-1}{\sqrt{3}}\right).\\
\end{split}
\end{equation}

We conclude that the topology of the Fermi surface in the $\Pi$-flux phase is not equivalent to the remaining four Abelian phases described so far. Indeed, the number of Dirac  fermions is  distinct, which reveals the existence of processes, deeply rooted in the non-Abelian phase, where  fermions scatter and pair-production (annihilation) is allowed. Such processes shall be discussed in detail in the forthcoming section, where a topological argument will clarify why such interacting events occur even when the Fermi gas is non-interacting.

\subsection{Effective low energy theories in a non-Abelian regime}

In order to identify the effects of the non-Abelian fields, we focus in the regime $\alpha=\beta\ll1$, where the topology of the Fermi surface can be worked out analytically. We note that a variety of different and interesting non-Abelian phases also occur for different fluxes $(\alpha,\beta)$, leaving their numerical treatment for the following sections. The energy band structure  
\begin{equation}
\label{non_Abelian_bands}
E(\textbf{k})=\pm t\sqrt{f_{\alpha,\textbf{k}}\pm\sqrt{g_{\alpha,\textbf{k}}}},
\end{equation}
can be expressed in terms of the following functions 
 \begin{equation}
 \begin{split}
 f_{\alpha,\textbf{k}}=&3+2\alpha^2+4\cos\textstyle{\frac{3k_x}{2}}\cos\textstyle{\frac{\sqrt{3}k_y}{2}}+2\cos\sqrt{3}k_y,\\
 g_{\alpha,\textbf{k}}=&-2\textstyle{\alpha^2}\left(2\cos3k_x\cos\textstyle{\sqrt{3}}k_y+(2+\alpha^2)\cos\textstyle{2\sqrt{3}}k_y\right.\\
 &\left.-4-\alpha^2-16\cos\textstyle{\frac{3k_x}{2}}\cos(\textstyle{\frac{\sqrt{3}k_y}{2}})\sin^{2}\textstyle{\frac{\sqrt{3}k_y}{2}}\right).\\
 \end{split}
 \end{equation}
 These energy bands have been represented in Fig.\ref{fig2}(c) for the particular case of $\alpha=0.2$, where one readily observes that the non-Abelian features of the external fields dramatically modify the  corresponding conical singularities. The valence and conduction bands each split into two components, and the conical singularities are displaced from the corners of the Brillouin zone (see also Fig.~\ref{fig3}(k) and \ref{fig3}(l)). However, we note that the two initial (two-fold degenerate) Dirac points give birth to four (non-degenerate) independent Dirac cones, so that $N_d=4$ remains constant in this non-Abelian regime. The exact position of the displaced Dirac points depends on the non-Abelian flux $\alpha$ as dictated by the following expressions
 \begin{equation}
 \begin{split}
 \textbf{K}_1&=\textstyle{ \left(0,\frac{2}{\sqrt{3}a}\arccos\left(\frac{-1+\sqrt{\alpha^2+2\alpha^4}}{2(1+\alpha^2)}\right)\right) },\\
  \textbf{K}_2&=\textstyle{ \left(-\frac{2\pi}{3a},\frac{2}{\sqrt{3}a}\arccos\left(\frac{1+\sqrt{\alpha^2+2\alpha^4}}{2(1+\alpha^2)}\right)\right) },\\
   \textbf{K}_3&=\textstyle{ \left(-\frac{2\pi}{3a},\frac{-2}{\sqrt{3}a}\arccos\left(\frac{1+\sqrt{\alpha^2+2\alpha^4}}{2(1+\alpha^2)}\right)\right) },\\
    \textbf{K}_4&=\textstyle{ \left(0,\frac{-2}{\sqrt{3}a}\arccos\left(\frac{-1+\sqrt{\alpha^2+2\alpha^4}}{2(1+\alpha^2)}\right)\right) }, \label{blabla}
    \end{split}
 \end{equation}
 which reduce to the Abelian points for $\alpha=0$ as expected. Once  the exact location of the non-Abelian Dirac points has been identified, it is possible to linearize the Hamiltonian in Eq.~\eqref{non_Abelian_momentum_hamiltonian} and obtain the continuum theory describing low-energy excitations. In the particular case of $\textbf{p}=\hbar\delta\textbf{k}=\hbar(\textbf{k}-\textbf{K}_1)$, we get the free massless Dirac Hamiltonian
 \begin{equation}
\label{Abelian_dirac_hamiltonian_bis}
H_{\text{D}}^{\textbf{K}}\approx c\left(\alpha_x\otimes\mathbb{I}_2\hspace{0.5ex}p_x+\alpha_y\otimes\mathbb{I}_2\hspace{0.5ex}p_y\right)+\Delta H,
\end{equation}
with a perturbation $\Delta H$ describing spin- and color-flip processes that arise due to the non-Abelian fluxes
\begin{equation}
\Delta H=\Delta_1\hspace{0.5ex}\sigma^+\otimes\mathbb{I}+\Delta_2\hspace{0.5ex}\sigma^+\otimes\tau^++\Delta_3\hspace{0.5ex}\sigma^+\otimes\tau^-+\text{h.c.}
\label{delta}
\end{equation}
Here, we have also defined
\begin{equation}
\begin{split}
\Delta_1&=\textstyle{(1+2\xi)-(\sqrt{3(1-\xi^2)}-\frac{\sqrt{3}}{2})\delta k_y+\ii(1-\xi)\delta k_x},\\
\Delta_2&=\textstyle{\alpha\sqrt{2}\eta_++\frac{\sqrt{3}\alpha}{\sqrt{2}}\eta_-\delta k_y}+\textstyle{\frac{\alpha}{\sqrt{2}}\eta_+\delta k_x},\\
\Delta_3&=\textstyle{-\alpha\sqrt{2}\eta_-+\frac{\sqrt{3}\alpha}{\sqrt{2}}\eta_+\delta k_y}+\textstyle{\frac{\alpha}{\sqrt{2}}\eta_-\delta k_x},\\
\end{split}
\end{equation}
with the following parameters  $\eta_{\pm}=(\xi\pm\sqrt{1-\xi^2})\ee^{\ii \pi/4}$, and $\xi=(-1+\sqrt{\alpha^2+2\alpha^4})/2(1+\alpha^2)$. From the expressions above, one can check that in the vanishing flux limit $\alpha\to0$, the perturbation also vanishes $\Delta H\to0$, and we recover the Abelian graphene theory. Let us remark here that the non-Abelian perturbation is essential to account for the displacement of the Dirac points. As follows from Fig.~\ref{low_energy_graphene_bands}(b), the dispersion relation around the singular points is not simply conical. Indeed, at low energies, the spectrum becomes linear in one direction, and displays a saddle point in the transverse direction. In this respect, we can expect a dramatic modification of the massless fermions properties due to their coupling to these exotic non-Abelian fields.

We conclude that the effect of the synthetic gauge fluxes is two-fold: 
they modify the Fermi surface by  transporting the Dirac points according 
to Eq. \eqref{blabla}, and they are also responsible of a different low-energy theory 
in Eq. \eqref{Abelian_dirac_hamiltonian_bis}. Indeed, the specific form of the perturbation in Eq. \eqref{delta} 
suggests the existence of certain emerging gauge fields that couple to the 
bare  massless fermions and give rise to this two-fold effect. A similar effect was explicitly shown in graphene-like structures with a Kekule distortion~\cite{kekule_graphene_1,kekule_graphene_2}.

\section{Topological quantum phase transitions}
\label{tqpt}

In the previous sections, we have thoroughly described the pattern of Dirac points and emerging relativistic fermions for different Abelian and non-Abelian phases. These phases differ by the total number of Dirac fermions and their distribution within the Brillouin zone. Below, we characterize the topological properties associated to the pattern of massless fermions by means of a  topological charge (i.e. topological invariant) $\mu=\pm1$ defined for each conical singularity~\cite{wen_zee}. Moreover,  the topological properties of the Fermi surface for non-interacting Fermi systems fully determine the underlying {\it quantum order}, the paradigm of order in quantum states beyond the Landau symmetry-breaking description~\cite{wen_book,volovik}. In this respect, the pattern of topological charges classifies  the  phases introduced above within different quantum-ordered universality classes having the same symmetry. 

We also study how a modification the gauge fluxes $(\alpha,\beta)$ connects different quantum orders (i.e. different  patterns of topological charges). In this respect, non-Abelian gauge fields induce a quantum phase transition (QPT) (i.e. a phase transition at $T=0$ driven by quantum fluctuations) between different vacua with the same symmetry, and cannot be  classified by the Landau symmetry-breaking paradigm. This class of QPTs have been previously studied in gapped systems such as the fractional QHE~\cite{wen_fqhe}, or gapless strongly correlated systems, such as certain types of spin liquids~\cite{wen_spin_liquid,wen_zee_spin_liquid}.  In the sections below, we show how such QPT occurs  in the realm of Fermi gases subjected to non-Abelian fields, which can be  probed in a cold atom experiment.

\subsection{Topological characterization: winding numbers}

The topology of the Fermi surface determines the type of effective theory and emergent symmetries at low energies. In order to classify the different universality classes in Fermi systems with gapless excitations, a topological winding number can be defined for the general Hamiltonian $H=\sum_{\textbf{k}}\Psi_{\textbf{k}}^{\dagger}H_{\textbf{k}}\Psi_{\textbf{k}}$, where $H_{\textbf{k}}$ was defined in Eq.~\eqref{non_Abelian_momentum_hamiltonian}. This Hamiltonian fulfills a particle-hole symmetry $\{\Gamma,H_{\textbf{k}}\}=0$, where $\Gamma=\text{diag}(\mathbb{I}_2,-\mathbb{I}_2)$, and allows us to define the following winding number
\begin{equation}
\nu=\frac{1}{4\pi\ii}\oint_{C}\text{tr}\left(\Gamma H_{\textbf{k}}^{-1}dH_{\textbf{k}}\right),
\end{equation} 
where $C$ is a curve defined in momentum space where the Hamiltonian $H_{\textbf{k}}$ presents no zeros~\cite{wen_zee}. Such a winding number, or the equivalent version
\begin{equation}
\mu=\frac{1}{2\pi\ii}\oint_{C}d\textbf{l}(\text{det}h_{\textbf{k}})^{-1}\partial_{\textbf{l}}\text{det}h_{\textbf{k}},
\end{equation} 
where $h_{\textbf{k}}=-t\sum_{j}\ee^{-\ii\textbf{k}\textbf{u}_j}U_{j}^{\dagger}$, has three important properties:

\vspace{0.5ex}
{\bf Dirac fermions:} The zeros of $H_{\textbf{k}}$, and thus conical singularities, are detected  inside the region delimited by the curve $C$, by means of a non-vanishing winding number $\mu\neq 0$.

\vspace{0.5ex}
{\bf Homotopic invariant:} The winding number is a homotopic invariant, and is thus independent with respect to distortions of the loop $C$ as far as $H_{\textbf{k}}^{-1}$ exists.

\vspace{0.5ex}
{\bf Topological invariant} The winding number is a topological invariant, and is thus independent with respect to local perturbations of the Hamiltonian $H_{\textbf{k}}\rightarrow H_{\textbf{k}}+\delta H_{\textbf{k}}$.
\vspace{0.5ex}

At half-filling, the Fermi surface is reduced to a set of points and we can characterize its topology by the pattern of conical singularities and their corresponding topological charges $(\textbf{K}_{j},\mu_j)$. As discussed in the next section, varying the gauge fluxes $(\alpha,\beta)$ along a non-Abelian path $\gamma:(\alpha_0,\beta_0)\rightarrow(\alpha_f,\beta_f)$, such that $|W|_{\gamma}\neq 2$, we observe that different quantum orders are connected by quantum phase transitions. We have  first determined the pattern of topological charges in each phase, and the corresponding universality classes:

\vspace{1ex}
{\bf Graphene universality class:} The quantum order in this equivalence class is determined by two positive topological charges $\mu=+1$, and two  negative topological charges $\mu=-1$ within the first Brillouin zone (see Fig.~\ref{fig3}(a)). There are thus $N_{d}=4$ Dirac points, which can be distributed in different positions in momentum space (see figs.~\ref{fig3}(a)-\ref{fig3}(h)), maintaining the same quantum numbers and thus belonging to the same universality class. We can thus conclude that all the graphene points $P_1,P_2,P_3,P_4$  describe the same quantum order.

\vspace{1ex}
{\bf $\Pi$-flux universality class:} The quantum order in this equivalence class is determined by four  positive topological charges $\mu=+1$, and four  negative topological charges $\mu=-1$ within the first Brillouin zone (see Fig.~\ref{fig3}(i)). Therefore, a total of $N_{d}=8$ species of gapless fermions arise.

\vspace{1ex}

As discussed so far, the different quantum orders are all determined by a distinct even number of gapless fermions. The  adiabatic variation of  the external gauge fluxes will give rise to processes, where the number of massless fermions is modified. Note however, that the total number of massless excitations shall always remain even, a distinctive feature of all the phases discussed above. A different prescription occurs for topological insulators with time-reversal invariance, where the number of massless excitations can be generally odd ~\cite{qshi_experiment,3d_ti_experiment,3d_ti_experiment2, top_insulator_nv,qsh_insulator_graphene,qsh_insulator_z2, top_insulator_2, qshi_semicond,3d_top_insulator,3d_top_insulator_3}. We believe that it would be interesting to study this type of topological phase transitions in such systems.

\subsection{Abelian path: absence of a quantum phase transition}

\begin{figure}[!hbp]
\centering
\begin{overpic}[width=8.50cm]{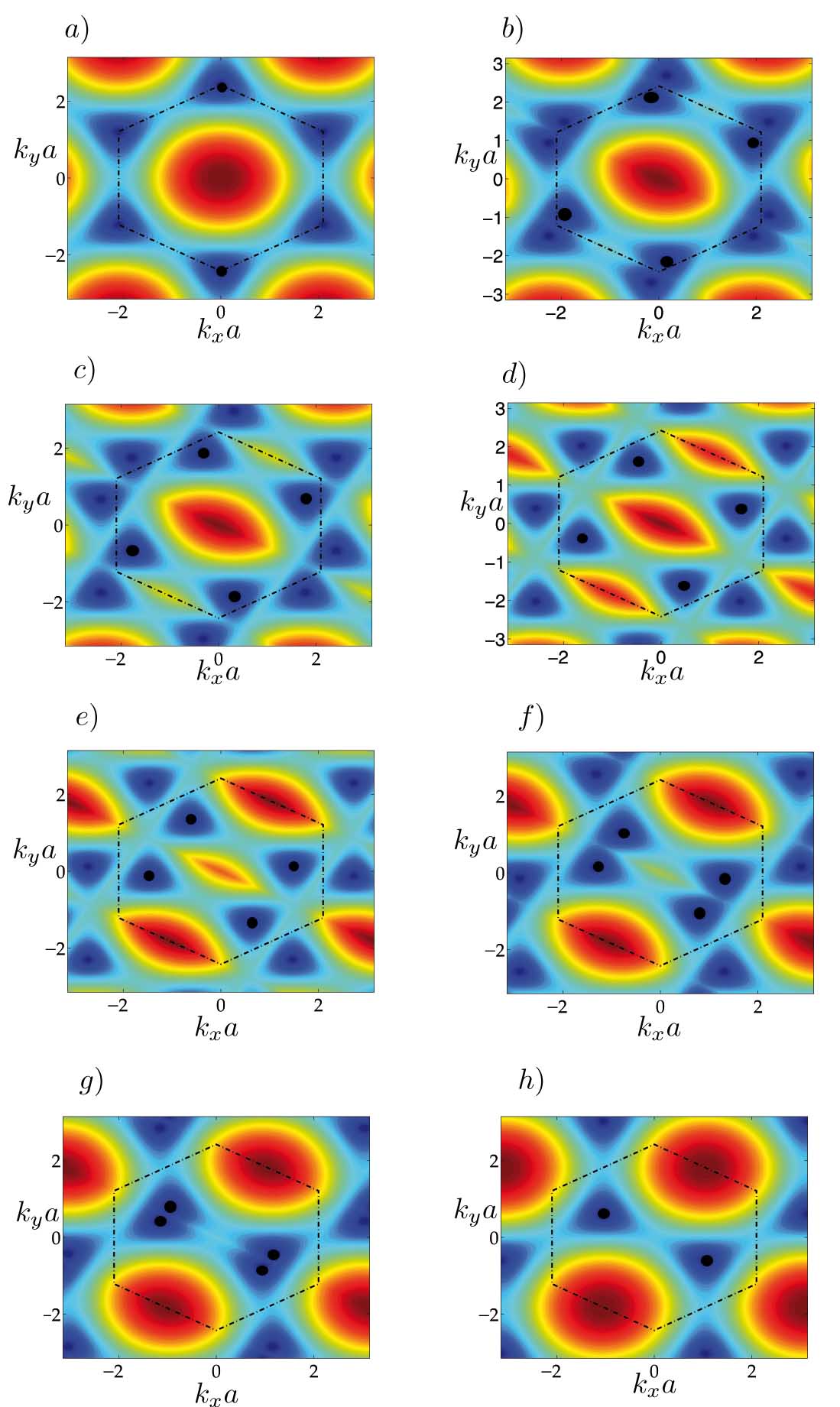}
\end{overpic}
\caption{ Pattern of Dirac points along the Abelian path connecting graphene $P_1$ to $4^{\text{th}}$ graphene point $P_4$. The gauge fluxes are the following (a) $\alpha=0,\beta=0$, (b) $\alpha=0.15\pi,\beta=0$, (c)$\alpha=0.30\pi,\beta=0$,  (d) $\alpha=0.45\pi,\beta=0$,  (e) $\alpha=0.60\pi,\beta=0$,  (f) $\alpha=0.75\pi,\beta=0$,   (g) $\alpha=0.90\pi,\beta=0$,   (h) $\alpha=\pi,\beta=0$.}
\label{fig19}
\end{figure}

In this section, we study the Fermi surface topology  as the gauge fluxes are modified along an Abelian path  $\gamma:(\alpha_0,\beta_0)\rightarrow(\alpha_f,\beta_f)$ such that $|W|_{\gamma}=2$. According to the previous section, this path only joins regimes belonging to the same universality class, that of graphene. Therefore, no  phase transition  occurs, and modifying the gauge flux only  transports the  $N_{d}=4$  topological charges in momentum space. We focus on the following Abelian path $\gamma_1(t)=(t\pi,0)$, with $t\in[0,1]$, connecting the $1^{\text{st}}$ graphene point $P_1$ to the $4^{\text{th}}$ graphene point $P_4$, as represented in Fig.~\ref{wilson_loop}. In figs.~\ref{fig19}(a)-\ref{fig19}(h), the distribution of conical singularities  along the Abelian path $\gamma_1$ is represented. 

In these figs.~\ref{fig19}(a)-\ref{fig19}(h), the Brillouin zone has been depicted by a dashed hexagon, and the conical singularities correspond to the dark-blue points where the valence and conduction bands touch. This set of contour figures represent the conduction band energy obtained by diagonalizing the Hamiltonian in Eq.~\eqref{non_Abelian_hopping} for different fluxes along the path $\gamma_1$.
  It clearly follows that, starting from graphene $P_1$ in Fig.~\ref{fig19}(a), the initial doubly-degenerate topological charges are split into a pair of Dirac points that travel in opposite directions (see Fig.~\ref{fig19}(b)). Note however that the total number of Dirac points $N_d=4$ and  topological charges $\mu=\{-1,-1,+1,+1\}$ are always preserved  from Fig.\ref{fig19}(a) to Fig.\ref{fig19}(h). Having the same pattern of topological charges, the different regimes along the Abelian path all belong to the same universality class, and have the same quantum order. The effect of a varying gauge flux along this path is simply the transport of topological charges in momentum space. Therefore, no  QPT connecting two different quantum orders exists, a fact that  also holds for any other Abelian path (e.g. that joining $P_2$ to $P_3$ in Fig.~\ref{wilson_loop}).

\subsection{Non-Abelian path: topological quantum phase transition}

From the previous section we conclude that the Abelian regime must be abandoned if different quantum orders are to be connected. Below, we show that a non-Abelian flux quench provides a rich scenario where scattering processes occur, involving the creation or the annihilation of massless Dirac fermions. Moreover, the pattern of topological charges is modified accordingly. Let us emphasize the exotic nature of these interaction processes, which  occur in a non-interacting gas. As shown below, only scattering of fermions with opposite topological charges is allowed. Moreover, we can draw an interesting analogy for two types of topological QPTs, which can be driven by a stimulated or spontaneous creation-annihilation of opposite topological charges.

\vspace{1ex}
{\bf Stimulated creation-annihilation of Dirac fermions}: 
Here, we study a non-Abelian path $\gamma_2$ that joins the $1^{\text{st}}$ graphene point $P_1$ to the $\Pi$-flux regime $P_5$, and then to the $3^{\text{rd}}$ graphene point $P_3$ (i.e. $\gamma_2(t)=(t\pi,t\pi)$, with $t\in[0,1]$, as represented in Fig.~\ref{wilson_loop}). 
This path is clearly non-Abelian, since $|W|_{\gamma_2}\neq2$ except at $t=\half$ where the $\Pi$-flux regime is reached, and, as  shown below, the pattern of topological charges determining the underlying  universality class is abruptly modified for certain critical fluxes $(\alpha_c,\beta_c)$. 

\begin{figure}[!hbp]
\centering
\begin{overpic}[width=8.50cm]{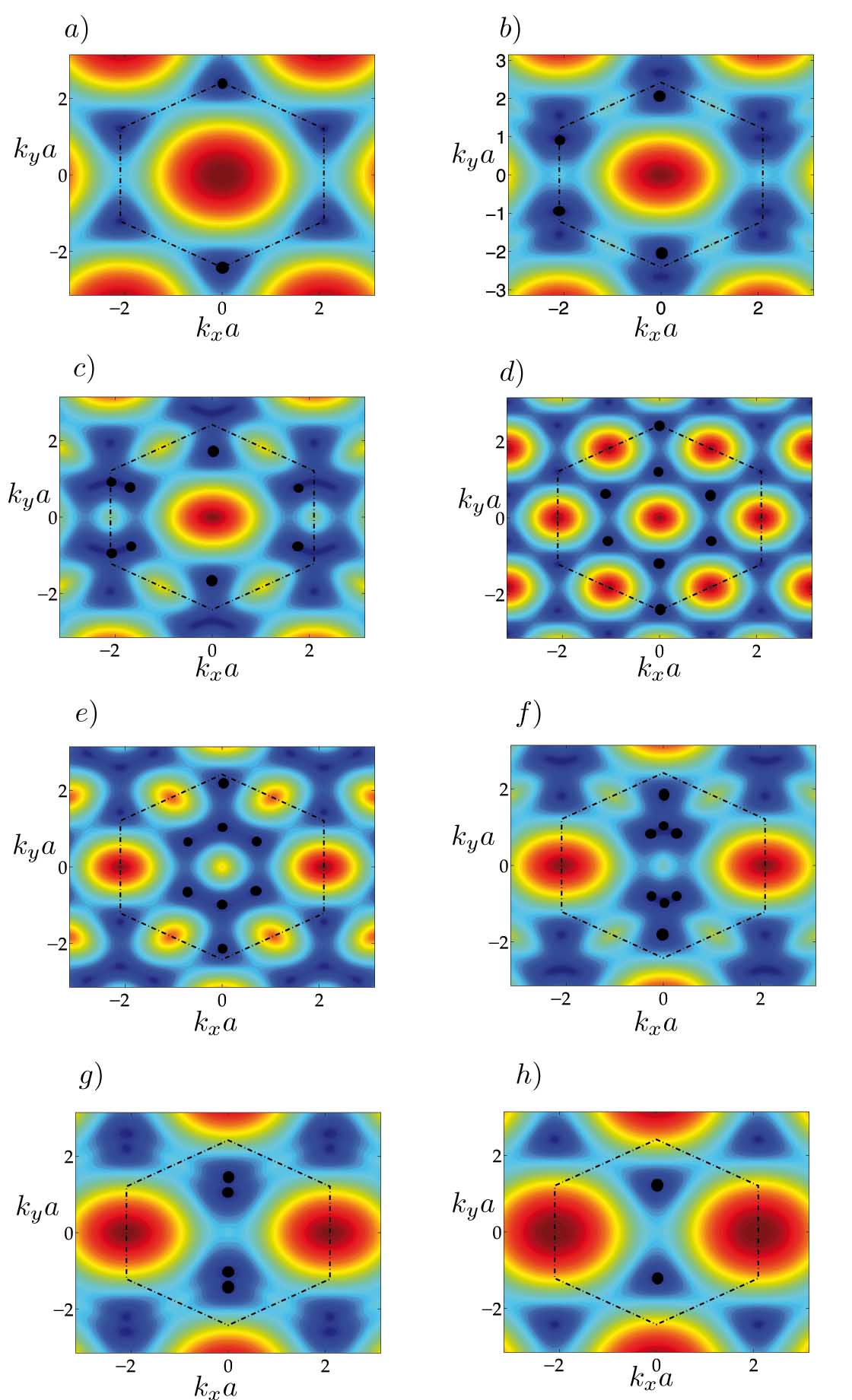}
\end{overpic}

\caption{ Pattern of Dirac points along the Abelian path connecting graphene $P_1$ to the $3^{\text{rd}}$ graphene point  $P_3$, passing through the $\Pi$-flux regime $P_5$. The gauge fluxes are the following (a) $\alpha=0=\beta$, (b) $\alpha=0.15\pi=\beta$, (c)$\alpha=0.30\pi=\beta$,  (d) $\alpha=0.50\pi=\beta$,  (e) $\alpha=0.60\pi=\beta$,  (f) $\alpha=0.75\pi=\beta$,   (g) $\alpha=0.90\pi=\beta$,   (h) $\alpha=\pi=\beta$.}
\label{fig21}
\end{figure}
\begin{figure}[!hbp]
\centering
\begin{overpic}[width=8.50cm]{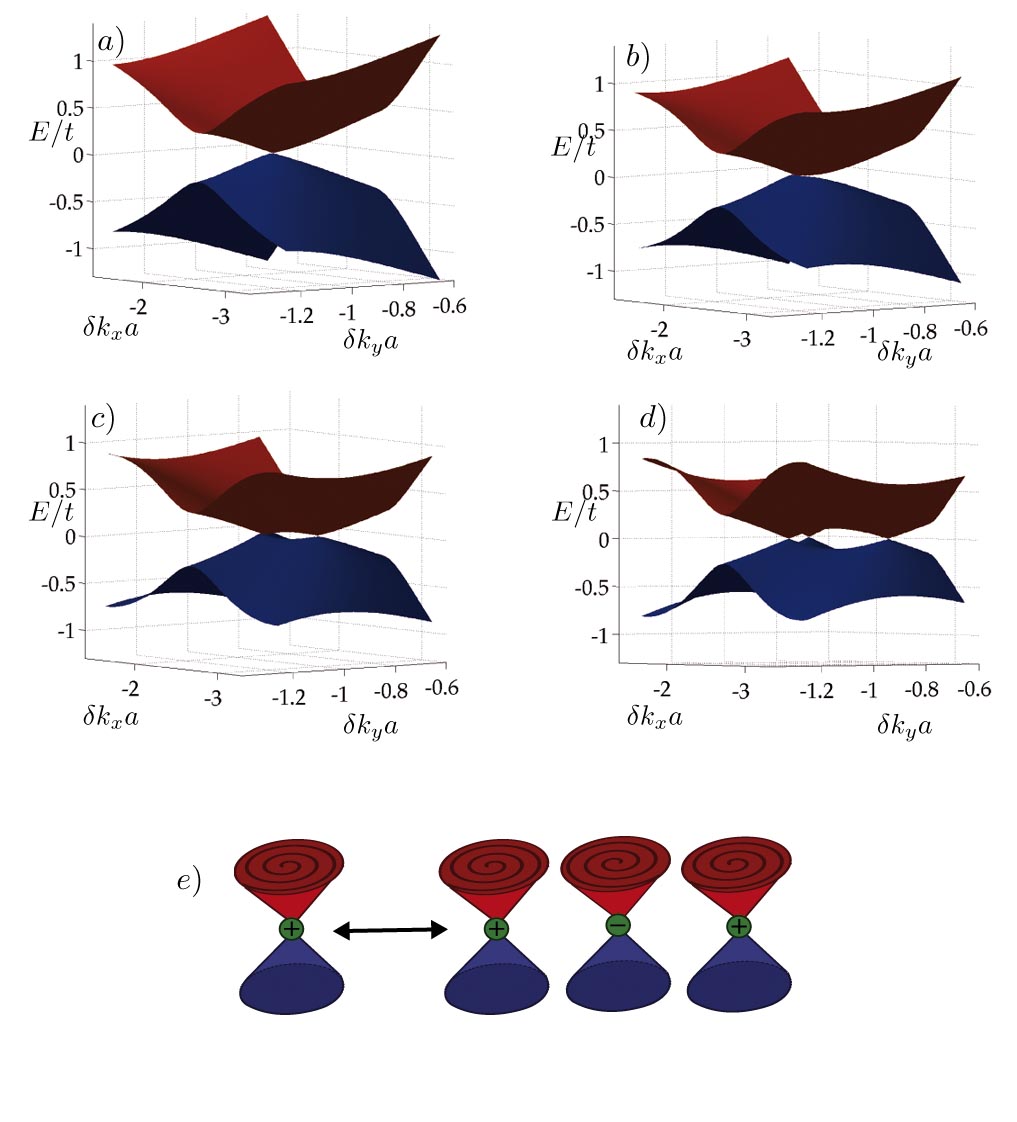}
\end{overpic}
\caption{ Energy bands dispersion around the \textbf{k}-space region where an induced pair production takes place. The gauge fluxes are the following (a) $\alpha=\beta=0.2\pi$, (b) $\alpha=\beta=0.25\pi$, (c)$\alpha=\beta=0.3\pi$,  (d) $\alpha=\beta=0.35\pi$. ( e) Scheme of the process of induced pair production representing the phenomena in Fig.~\ref{fig21}(c), and the reverse annihilation process (time-reversed event) corresponding to Fig.~\ref{fig21}(f).}
\label{fig22}
\end{figure}

In figs.~\ref{fig21}(a)-\ref{fig21}(h), we observe how the initial doubly-degenerate Dirac points (Fig.~\ref{fig21}(a)) are split into a pair of fermions which move along the vertical axis in opposite directions (Fig.\ref{fig21}(b)). Remarkably, in Fig.~\ref{fig21}(c) we can observe how additional  Dirac fermions are being created along the borders of the Brillouin zone.  A closer inspection of the energy band dispersion around $\textbf{k}=\frac{2\pi}{3a}(-1,\frac{-1}{\sqrt{3}})$ (i.e. the region where these gapless fermions are created), reveals that there is a process of induced or stimulated pair production (see figs.~\ref {fig22}(a)-\ref{fig22}(d)), where a couple of massless fermions is created in the presence of a third gapless mode. Interestingly enough, the pair of created fermions carry opposite topological charges $\mu=\pm1$, so that the total winding number is conserved during the scattering process (see also Fig.\ref{fig22}(e)). Note that the production of fermionic gapless modes occurs at $T=0$,  and thus a QPT between different quantum orders takes place. In the same manner symmetry-breaking phase transitions lead to gapless Goldstone bosons, QPTs driven by fluctuations of quantum order can give rise to gapless fermionic excitations. Also, analogous to the symmetry protection of Goldstone modes, these massless fermions are protected by quantum order (i.e. topological invariant charges $\mu=\pm 1$). 

The quantum phase transition taking place around $\alpha_c=\beta_c=\frac{\pi}{4}$, connects two different phases with $N_d=4$ and $N_d=8$ low energy Dirac points, so that the total topological charge is conserved. Let us also note that there is a critical point at $\alpha_c=\beta_c=\frac{3\pi}{4}$ where the inverse process takes place. In this case, two massless fermions with opposite charges are annihilated and connect the phases with $N_d=8$ and $N_d=4$ fermions. It is also interesting to remark that at criticality $\alpha_c=\beta_c=\frac{\pi}{4}$, one can show that the massless excitations are highly anisotropic, being relativistic in one direction, and non-relativistic along the transverse axis.

\begin{figure}[!hbp]
\centering
\begin{overpic}[width=8.50cm]{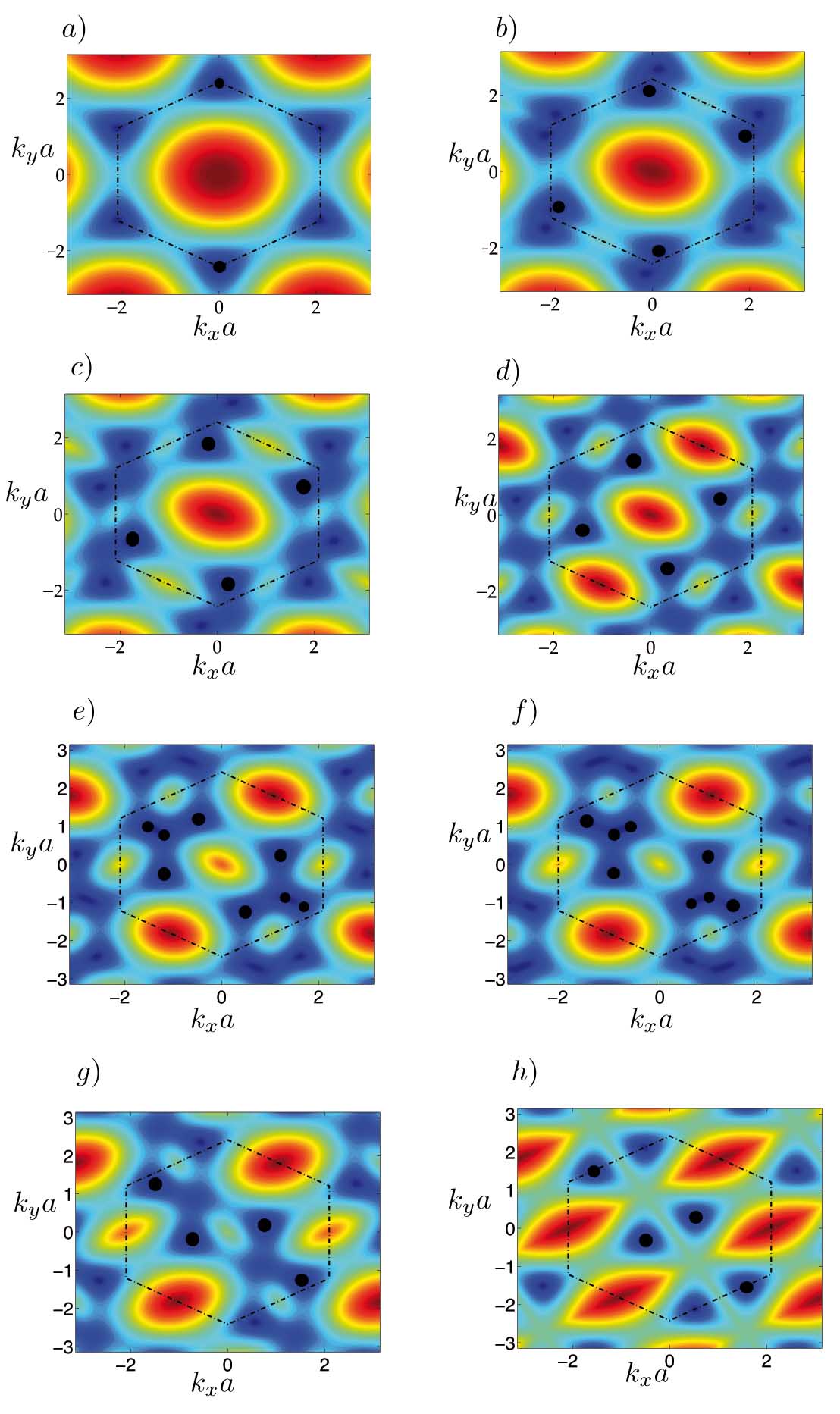}
\end{overpic}

\caption{ Pattern of Dirac points along the non-Abelian path connecting graphene $P_1$ to $\alpha=\pi,\beta=\half\pi$. The gauge fluxes are the following $\beta=\half\alpha$ (a) $\alpha=0$, (b) $\alpha=0.15\pi$, (c)$\alpha=0.3\pi$,  (d) $\alpha=0.5\pi$,  (e) $\alpha=0.6\pi$,  (f) $\alpha=0.7\pi$,   (g) $\alpha=0.8\pi$,   (h) $\alpha=\pi$.}
\label{fig24}
\end{figure}

\vspace{1ex}
{\bf Spontaneous creation-annihilation of Dirac fermions:}  Below, we prove that spontaneous processes of pair creation-annihilation are  also possible, where the pair of oppositely charged fermions are created from a local vacuum (i.e. locally gapped spectrum). Here, we study a non-Abelian path $\gamma_3$ that joins the $1^{\text{st}}$ graphene point $P_1$ to the $(\alpha,\beta)=(\pi/2,\pi/4)$ (i.e. $\gamma_3(t)=(t\pi,t\pi/2)$, with $t\in[0,1]$). The different Dirac points along such a path have been represented in figs.~\ref{fig24}(a)-\ref{fig24}(h). In Fig.~\ref{fig24}(d), we observe how a pair of fermion-antifermion is spontaneously created from a locally gapped vacuum around two corners of the Brillouin zone $\textbf{k}=\frac{2\pi}{3a}(\mp1,\frac{\pm1}{\sqrt{3}})$ for $\alpha=2\beta=0.6\pi$. Then in Fig.~\ref{fig24}(e), we observe how one of the recently created fermions scatters with another Dirac point and how they are spontaneously annihilated for $\alpha=2\beta=0.7\pi$.

\begin{figure}[!hbp]
\centering
\begin{overpic}[width=8.500cm]{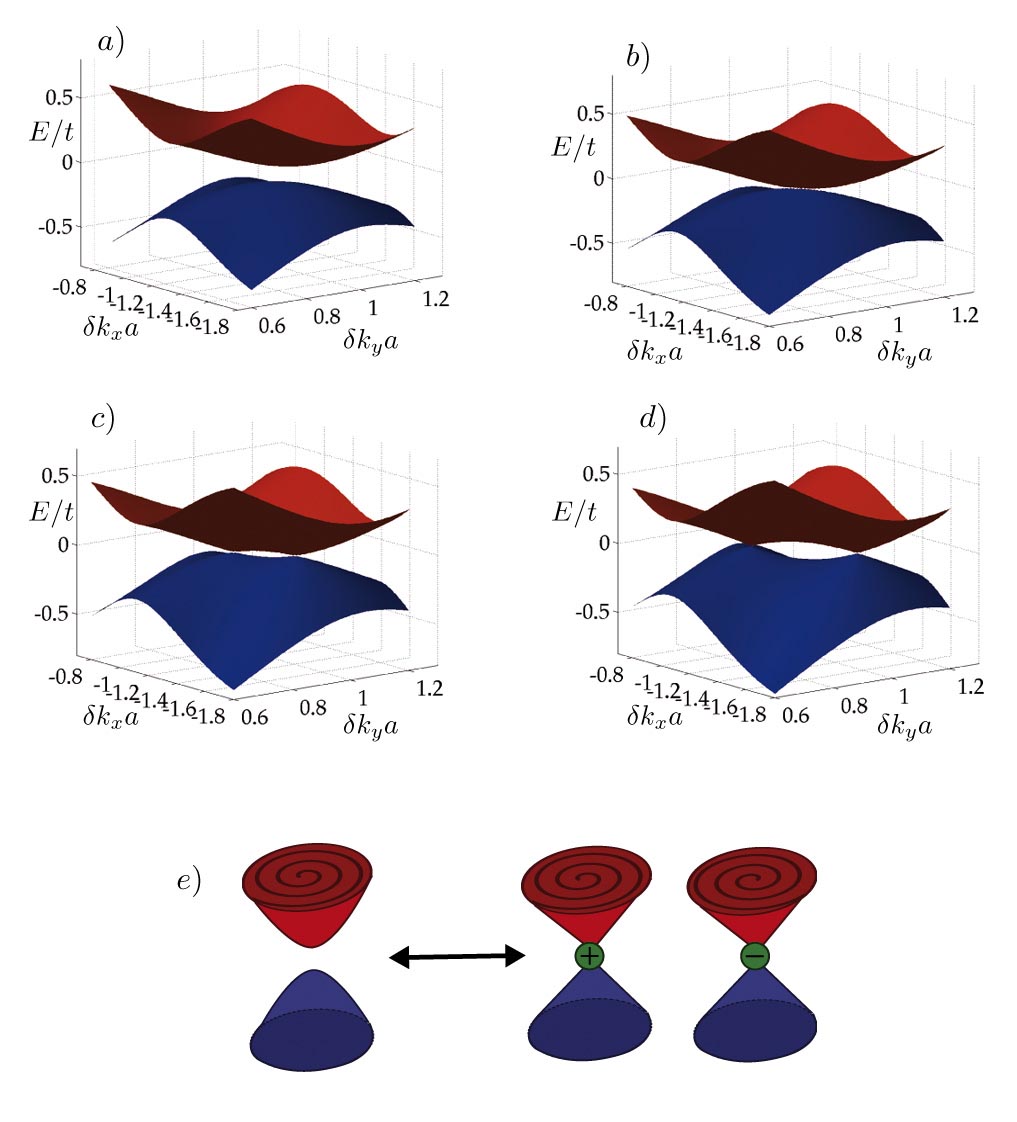}
\end{overpic}
\caption{ Energy bands dispersion around the \textbf{k}-space region where a spontaneous pair production takes place. The gauge fluxes are the following (a) $\alpha=2\beta=0.6\pi$, (b) $\alpha=2\beta=0.63\pi$, (c)$\alpha=\beta=0.64\pi$,  (d) $\alpha=2\beta=0.66\pi$. (e) Scheme of the process of spontaneous pair production, and annihilation (time-reversed event).}
\label{fig25}
\end{figure}

In figs.~\ref{fig25}(a)-\ref{fig25}(d) we study the energy band landscape as the gauge fluxes are swept across the critical point $\alpha_c=2\beta_c=0.63\pi$. Initially, the low energy excitations are clearly gapped and for low enough energies we have a local vacuum. Then, as the gauge fluxes increase, we observe how the local gap vanishes and two massless fermions appear. The topological charge distribution is schematically shown in Fig.~\ref{fig25}(d), where the total topological charge is once more conserved. Let us note that for  $\alpha_c=2\beta_c=0.7\pi$ (see Fig.~\ref{fig25}(f)), the time-reversed scattering process takes place. Here, we observe a pair fermion-antifermion being annihilated due to their opposite charges. These events of spontaneous pair creation or annihilation connect two distinct quantum orders $N_d=4\to 8$ by means of a non-symmetry breaking quantum phase transition. A similar spontaneous creation-annihilation process is also illustrated in fig.~\ref{QHE8} (see Sec.~\ref{QHEsect}).   In the following sections, we address the issue of characterizing the above topological phases through accessible measurement techniques in the optical-lattice setup.

Let us close this section by comparing the topological QPT presented in this work, to the semi-metal to insulator transition in bare graphene-like systems~\cite{duan_graphene_ol,graphene_DP_merging,garphene_DP_merging_2,graphene_DP_merging_strain,DP_merging_OL,DP_kz,DP_semi}. In graphene, a modulation of the hopping anisotropy  induced by strain can lead to the merging of Dirac points, which turns graphene into an insulator. An insulating gap can also be dynamically induced by quenching the sublattice energy imbalance~\cite{DP_semi}. Conversely, in the non-Abelian honeycomb lattice, the merging of Dirac points does not lead to a global gap, and the semi-metallic phase is always preserved. Accordingly, this optical-lattice setup offers a broader spectrum of topological QPTs, where both spontaneous and stimulated creation-annihilation of Dirac fermions occur.

\section{The quantum Hall measurements}
\label{QHEsect}

The presence of several distinct Dirac points within the first Brillouin zone leads to dramatic consequences in the physical properties of the system. In the low-energy regime, these relativistic effects are adequately described by the underlying effective Dirac theories. An interesting effect to be considered in this context is the so-called anomalous quantum Hall effect (AQHE)~\cite{jackiw_qhe,schakel_qhe}, which has been discovered in graphene sheets subjected to external magnetic fields~\cite{sharapov_qhe,guinea_qhe,grap_anomalous_qhe_exp1,grap_anomalous_qhe_exp2}. This effect is characterized by the peculiar sequence of plateaus in the transverse conductivity of the system as a function of the chemical potential: in the low-energy regime, where  the spectrum exhibits the so-called relativistic Landau levels~\cite{rel_landau_levels,rel_landau_levels_graph,grap_landau_levels_exp}, the Hall conductivity is quantized according to  $\sigma_{xy}=\pm\frac{4}{h}(\nu+\frac{1}{2})$, where $\nu$ is an integer, $h$ is Planck's constant and we set $e=1$. The Hall measurements show steps of four that can be traced back to the presence of two independent two-fold degenerate Dirac points $N_d=4$, and a half-integer anomaly due to the zero-energy mode associated to the self-conjugate lowest Landau level. 

In  previous sections, it has been shown how several Dirac points are induced in our cold-atom system. In particular, we have shown that the number of independent Dirac cones $N_d$ is not necessarily conserved as the non-Abelian fluxes $(\alpha , \beta)$ are varied. The aim of this section is to show that the number of emerging massless fermions $N_d$ can be inferred from Hall measurements. In particular, we show that the Hall plateaus at zero Fermi energy $E_F=0$ fulfill
\begin{equation}
\sigma_{xy}=\pm\frac{N_d}{2h},
\label{conjecture}
\end{equation}
and thus provide a Dirac fermion witness. Moreover, the robustness of this witness is associated to the system zero-modes, and is guaranteed by the Hamiltonian particle-hole symmetry and topological index theorem~\cite{jackiw_qhe}. The relation of the Hall conductivity to the number of Dirac fermions 
in Eq. \eqref{conjecture} can be conjectured from the theoretical results in \cite{jackiw_qhe,schakel_qhe,sharapov_qhe,guinea_qhe}. 
Below, we confirm this conjecture by numerically computing the Chern numbers 
associated to the energy band structure, and thus estimating the corresponding 
Hall conductivities. Furthermore, this conductivity can be 
experimentally  accessed in cold atoms  by the so called 
Streda formula \cite{Umu}. \\

 In this section, we shall first review how such measurements can  be performed in cold-atom experiments by analyzing the density profiles. Then, we show how the AQHE taking place in this system gives us crucial information on the topological QPTs that occur as the non-Abelian fluxes are varied. In particular, we discuss under what conditions the Hall measurements are indeed witnesses  of such transitions.

\subsection{The anomalous quantum Hall effect and the density profiles}

\begin{center} 
\begin{figure}
\begin{center}
\hspace{-0.1cm}{\scalebox{15}{\includegraphics{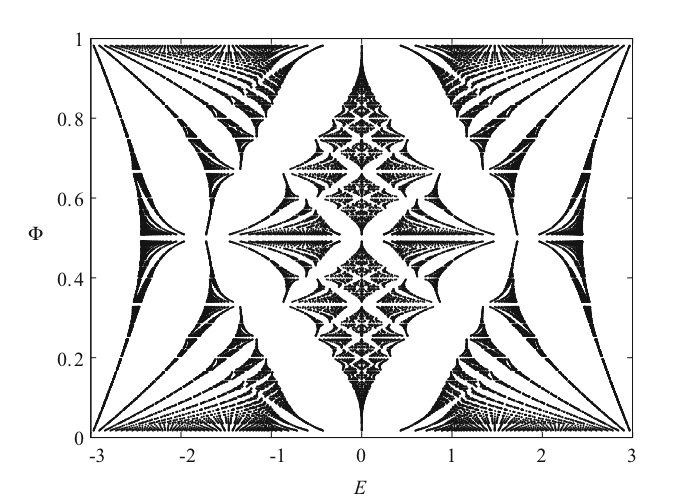}}}
\hspace{-0.1cm}{\scalebox{0.35}{\includegraphics{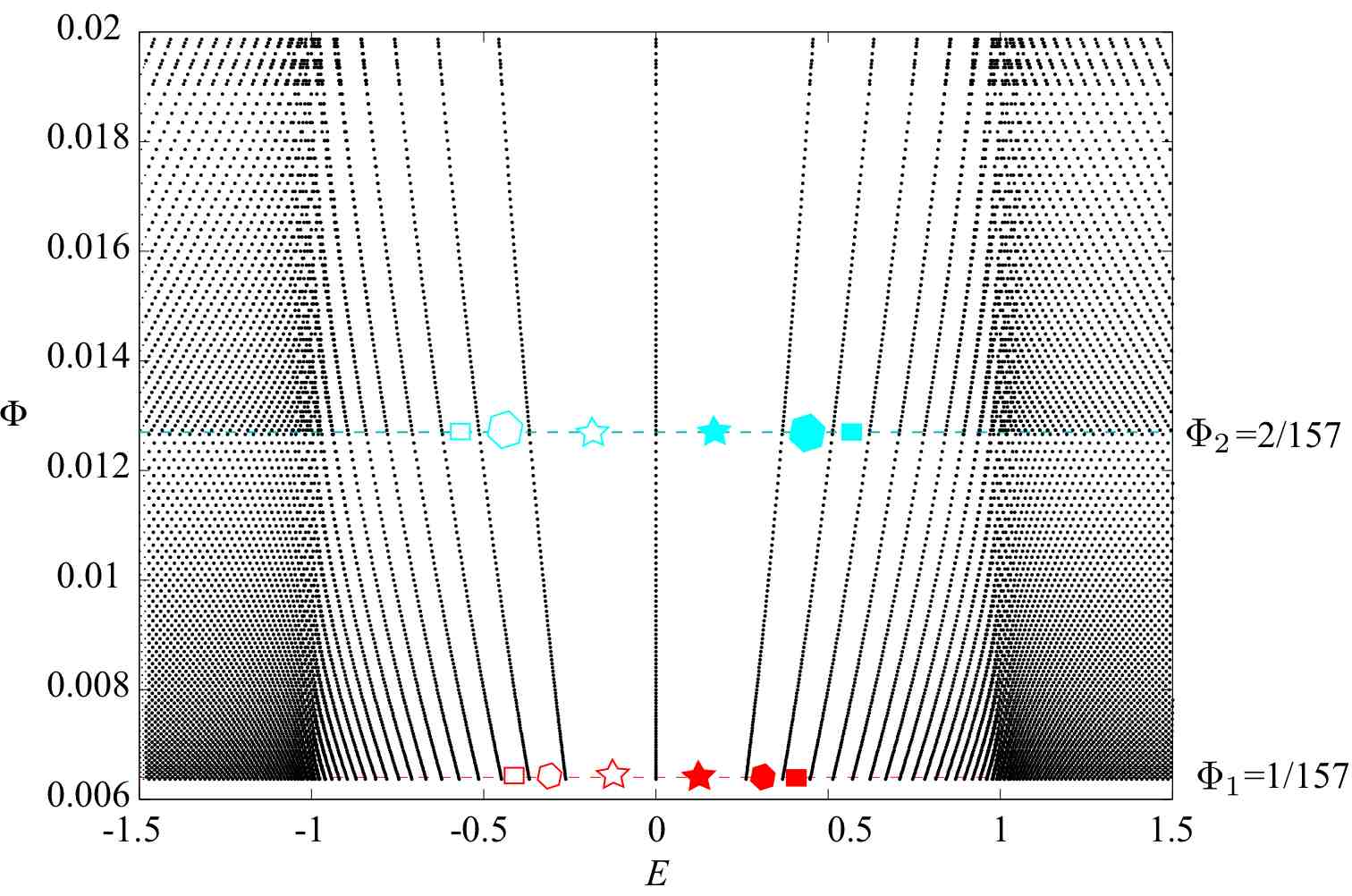}}}
\caption{\label{QHE1}  (top) The Hofstadter butterfly for an hexagonal lattice subjected to a magnetic flux $\Phi$. Here one sets $\alpha=\beta=0$. (bottom) The same butterfly spectrum in the vicinity of the Dirac regime $E \in [-1 , 1]$. The main energy gaps at $\Phi_1=1/157$ and $\Phi_2=2/157$ are labelled with different symbols.} 
\end{center}
\end{figure} 
\end{center}

In order to provoke the quantum Hall effect in our system, one has to open the central gap. This can be achieved by subjecting the system to an additional Abelian gauge field (i.e. an artificial magnetic field). Such synthetic magnetic fields can be produced for neutral atoms by rotating the lattice \cite{polini, polini2,Holland1,Holland2,Tung},  by considering laser-induced methods \cite{spielman,OL_Abelian_field,OL_Abelian_field_2,OL_Abelian_field_3,OL_ab_field_dark_states_1,OL_ab_field_dark_states_2}, or even by immersing the system in a rotating BEC \cite{imme}. The presence of an external Abelian gauge field modifies the hopping operators introduced in Sect. III, which now take the form
\begin{equation}
U_1=e^{i 2 \pi \Phi x} e^{i \alpha \sigma_x}, \qquad U_2=\mathbb{I}, \qquad U_3=e^{i \beta \sigma_y}
\end{equation} 
where $\Phi$ is the effective magnetic flux quanta per unit cell. A clear manifestation of the external magnetic field can already be found by computing the energy spectrum: when the flux is given by the ratio of two integers, namely $\Phi=p/q$, the initial band structure is split into several subbands. For such rational Abelian fluxes, we note that the Hamiltonian takes the form of a $4q \times 4 q$ matrix. Therefore the number of gaps highly depends on the integer $q$, but also on the specific values of the non-Abelian fluxes $\alpha$ and $\beta$. When plotted as a function of the flux $\Phi$, these numerous subbands form a fractal structure which reminds the well-known Hofstadter butterfly \cite{hofstadter}, as depicted in Fig. \ref{QHE1} for the Abelian case $\alpha=\beta=0$.  \\

\begin{center} 
\begin{figure}
\begin{center}
\hspace{-0.1cm}{\scalebox{0.19}{\includegraphics{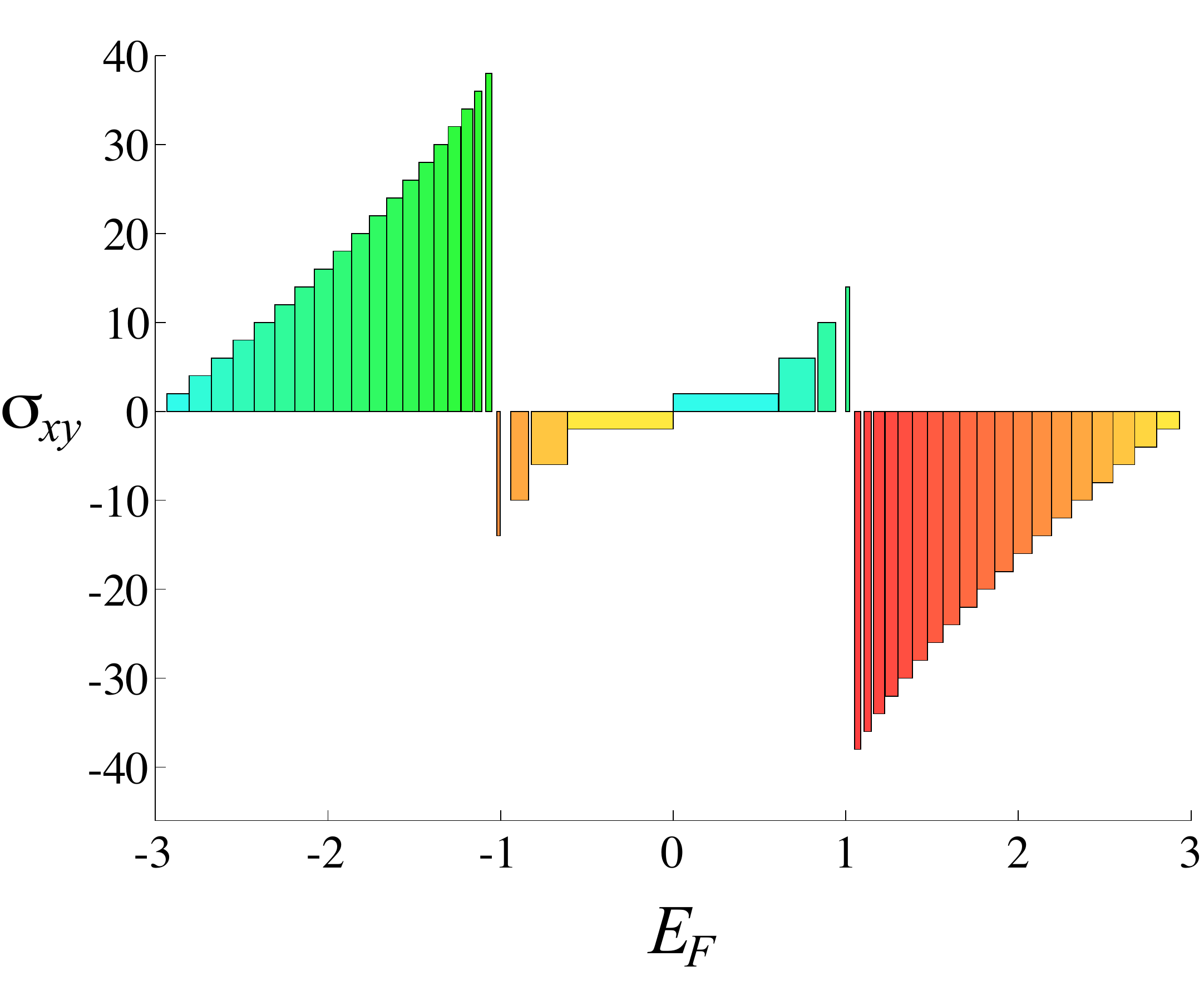}}}
\hspace{-0.1cm}{\scalebox{0.19}{\includegraphics{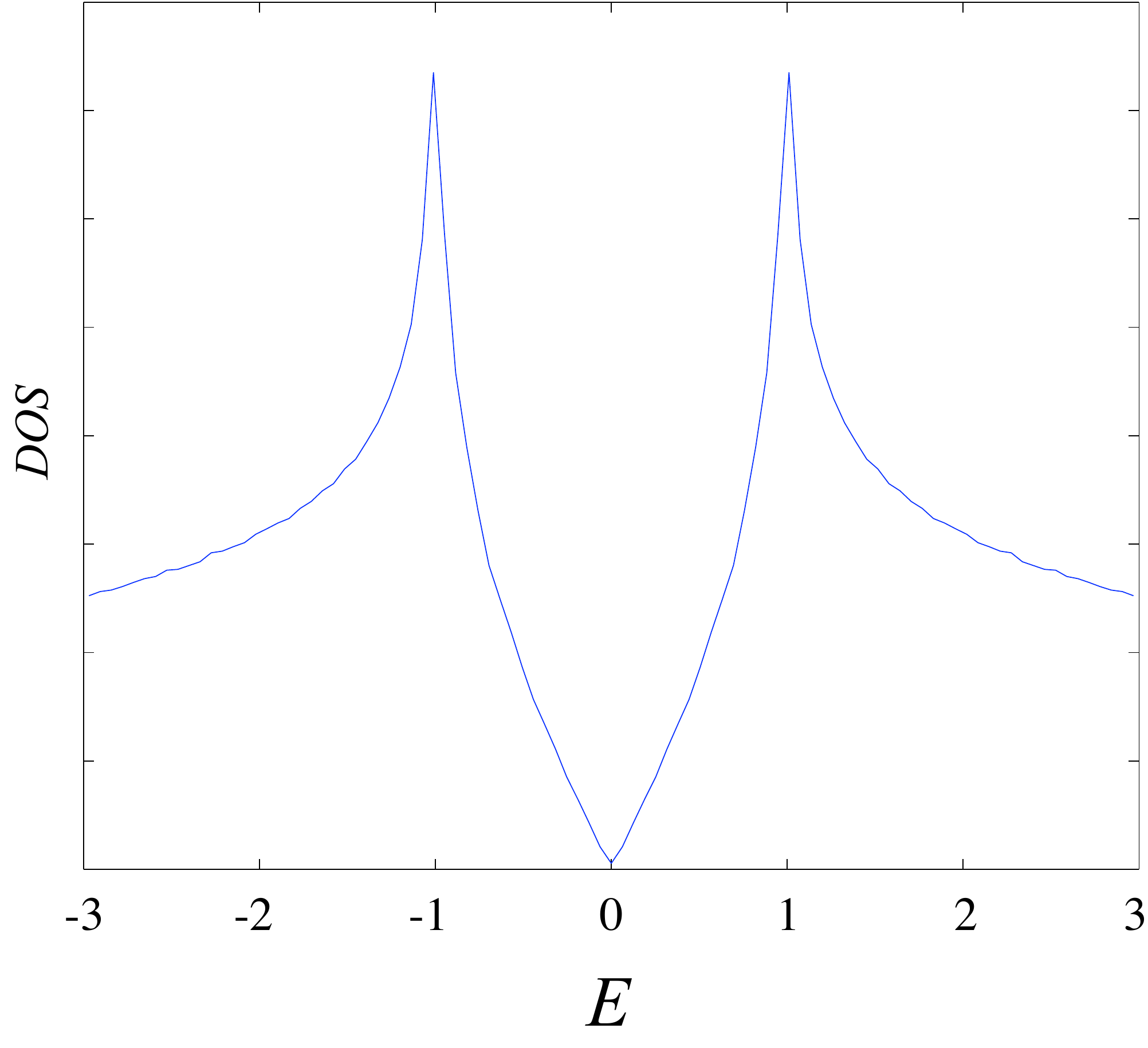}}}
\caption{\label{QHE2}  (a). Hall plateaus in the low-flux regime $\Phi \ll 1$  and $\alpha=\beta=0$ . (b) The density of states in the same configuration.} 
\end{center}
\end{figure} 
\end{center}

In the cold-atom framework, the conductivity tensor measures the response of the system to an external perturbation, such as a lattice acceleration, which generates a current. In the presence of an additional magnetic field, a Hall current is generated and this transverse transport is dictated by the Hall conductivity $\sigma_H$. When the Fermi energy $E_F$ lies in a spectral gap, the Hall conductivity can be expressed as a sum of topological invariants 
\begin{equation}
\sigma_H= \frac{1}{h} \sum_{E_n < E_F} N_{ch} (E_n),
\label{sigmaH}
\end{equation}
where $N_{ch} (E_n)$ is the Chern number associated to the filled energy band $E_n$, which  is necessarily an integer \cite{Kohmoto1985}. Consequently, the sum in Eq. \eqref{sigmaH} characterizes the different topological phases  associated to the spectral gaps. These quantum Hall phases can be reached by varying the Fermi energy $E_F$, namely, by controlling the atomic filling factor. The computation of the Chern numbers can be achieved numerically through diverse techniques \cite{nathan-book,OL_non_ab_qhe,Fukui} and leads to the complete description of the QHE in  our system. In Fig. \ref{QHE2} (a), we show the Hall plateaus  in the Abelian case $\alpha=\beta=0$, as recently discovered in graphene. One clearly distinguishes the Dirac regime, situated in the range $E_F \in [-1 ,1]$, where the anomalous double steps and a half-integer plateau are observed. It should be stressed  that the range of this relativistic regime, where the dispersion relation is linear, is known to be bounded by singularities in the density of states (see Fig. \ref{QHE2} (b)).   \\

\begin{center} 
\begin{figure}
\begin{center}
\vspace{0.cm}{\scalebox{0.3}{\includegraphics{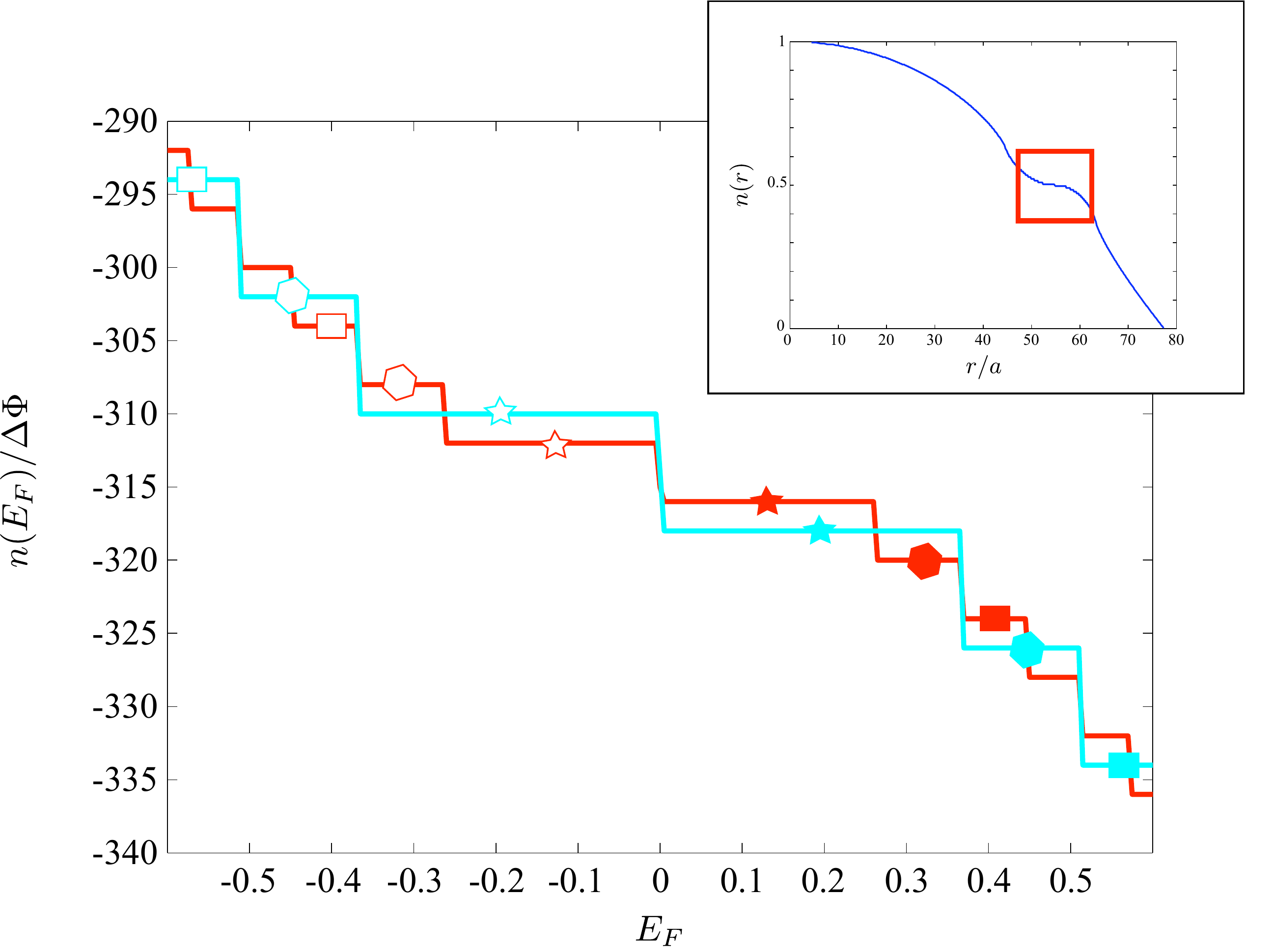}}}
\caption{\label{QHE3}  main figure: Density profiles $n(E_F)/\Delta \Phi$ in the Abelian case $\alpha=\beta=0$: for $\Phi_1=1/157=6.4 10^{-4}$ (dark red) and for $\Phi_2=2/157=12.7 10^{-4}$ (light cyan). Here the density $n(E_F)$ is divided by $\Delta \Phi= \Phi_1 - \Phi_2$, and the plateaus correspond to the energy gaps that are highlighted in Fig.~\ref{QHE1}(b) with the same symbols. Upper right: Density profile $n(r)$ in the presence of a typical harmonic potential for $\Phi_1=1/157$. The red rectangle specifies the relativistic regime which is considered in the main figure.} 
\end{center}
\end{figure} 
\end{center}

Although optical-lattice experiments involving atomic currents are on their way, it is preferable to exploit another method to evaluate the Hall conductivity. Indeed, several works have recently proposed that the quantized Hall conductivity could be measured through an elegant analysis of the density profiles \cite{Umu,fuxiang,shao,fuxiang2}. Furthermore, this efficient method takes into account the presence of a harmonic potential   $V_{\txt{harm}}(r)$ that generally confines the atoms within the optical lattice. If this external potential varies slowly compared to the lattice spacing, one may use the local-density approximation (LDA) and define a local Fermi energy as 
\begin{equation}
E_F(r)= E_F-V_{\txt{harm}}(r).
\label{LDA}
\end{equation}
 The atomic density is then simply obtained through the expression
$n (r)= \int d \epsilon \, D_{\textnormal{}}(\epsilon) \, \Theta \bigl ( E_{\textnormal{F}}(r) - \epsilon \bigr )$, where $D_{\textnormal{}}(\epsilon)$ is the density of states of the uniform system. Each time the Fermi energy $E_{\textnormal{F}}(r)$ falls in a gap, the density $n (r)$ depicts a plateau: there is a one-to-one correspondence between the spectral gaps represented in Fig. \ref{QHE1} and the plateaus of the density profiles $n(r)$. We illustrate this phenomenon in Fig. \ref{QHE3}, where two density profiles  for $\Phi_1=1/157$ and $\Phi_2=2/157$ are represented, and where the correspondence with the gaps  in Fig. \ref{QHE1} (b) has been highlighted with symbols. In this figure, we focus on the relativistic region  around $E_F=0$ and  set $\alpha=\beta=0$. Note that in the following, the density per plaquette is studied as a function of the Fermi energy $n(E_F)$, being  density profiles $n(r)$  simply obtained by considering the LDA relation in Eq.\eqref{LDA}. In the upper pannel of Fig. \ref{QHE3}, a typical density profile $n(r)$ obtained for $\Phi_1=1/157$ is shown: the relativistic regime can already be identified by an unusual behavior (see inside the red rectangle). \\

Having identified the unusual plateaus in the density profiles around $E_F$, one may now evaluate the Hall conductivity through the Streda formula \cite{Umu}, which takes the form
\begin{equation}
\sigma_H (E_F)= \frac{1}{h} \frac{\partial n}{\partial \Phi} \Big \vert_{E_F}.
\label{streda}
\end{equation}
First of all, let us show how the Streda relation \eqref{streda} allows one to detect the AQHE at the Graphene point $P_1$ ($\alpha=\beta=0$). Comparing the two density profiles  in Fig. \ref{QHE3}, the Hall conductivity inside the main gaps around $E_F=0$ becomes
\begin{align}
\sigma_H (\txt{empty rectangle})h&=\frac{n_1}{\Delta \Phi}-\frac{n_2}{\Delta \Phi} =-304+294=-10 \notag \\
\sigma_H (\txt{empty hexagon})h&=\frac{n_1}{\Delta \Phi}-\frac{n_2}{\Delta \Phi}= -308+302=-6 \notag \\
\sigma_H (\txt{empty star})h&=\frac{n_1}{\Delta \Phi}-\frac{n_2}{\Delta \Phi}= -312+310=-2 \notag \\
\sigma_H (\txt{filled star})h&=\frac{n_1}{\Delta \Phi}-\frac{n_2}{\Delta \Phi}=-316+318= 2 \notag \\
\sigma_H (\txt{filled hexagon})h&=\frac{n_1}{\Delta \Phi}-\frac{n_2}{\Delta \Phi}= -320+326=6 \notag \\
\sigma_H (\txt{filled rectangle})h&=\frac{n_1}{\Delta \Phi}-\frac{n_2}{\Delta \Phi}= -324+334=10 \notag
\end{align}
where $n_1(E_F)/\Delta \Phi $ [resp. $n_2(E_F)/\Delta \Phi$] corresponds to the density profile for $\Phi_1=1/157$ (dark red) [resp. $\Phi_2=2/157$ (light cyan)] in Fig. \ref{QHE3}, and $\Delta \Phi= \Phi_1 - \Phi_2$. Consequently, and in perfect agreement with the Hall plateaus illustrated in Fig. \ref{QHE2} (a), the Hall conductivity undergoes the sequence $\sigma_{H}h\in\{... , -10, -6, -2, 2, 6, 10, ...\}$, and therefore agrees with the relativistic prediction $\sigma_{H}=\pm\frac{4}{h}(\nu+\frac{1}{2})$. Moreover, this example shows how a precise analysis of the density profile measurements obtained for two close values of the flux $\Phi$ allows one to observe the AQHE proper to the relativistic regime. Furthermore, these density profiles constitute a powerful tool to determine the number of Dirac cones situated inside the first BZ, since the latter is directly related to the jumps observed in the anomalous Hall sequences $N_d=4$. This important result will be exploited in the following section to study how the density profiles are modified as one varies the non-Abelian fluxes $\alpha$ and $\beta$. At this point, it is worth noticing that the density profiles, and thus the Hall plateaus, are not affected as one travels along an Abelian path (see Sect. IV B). The latter fact is in agreement with the fact that $N_d=4$ remains constant along such a path and that no topological QPT is thus observed. \\

\subsection{The Hall plateaus along a non-Abelian path} 

\begin{center} 
\begin{figure}
\begin{center}
\vspace{0.cm}{\scalebox{4}{\includegraphics{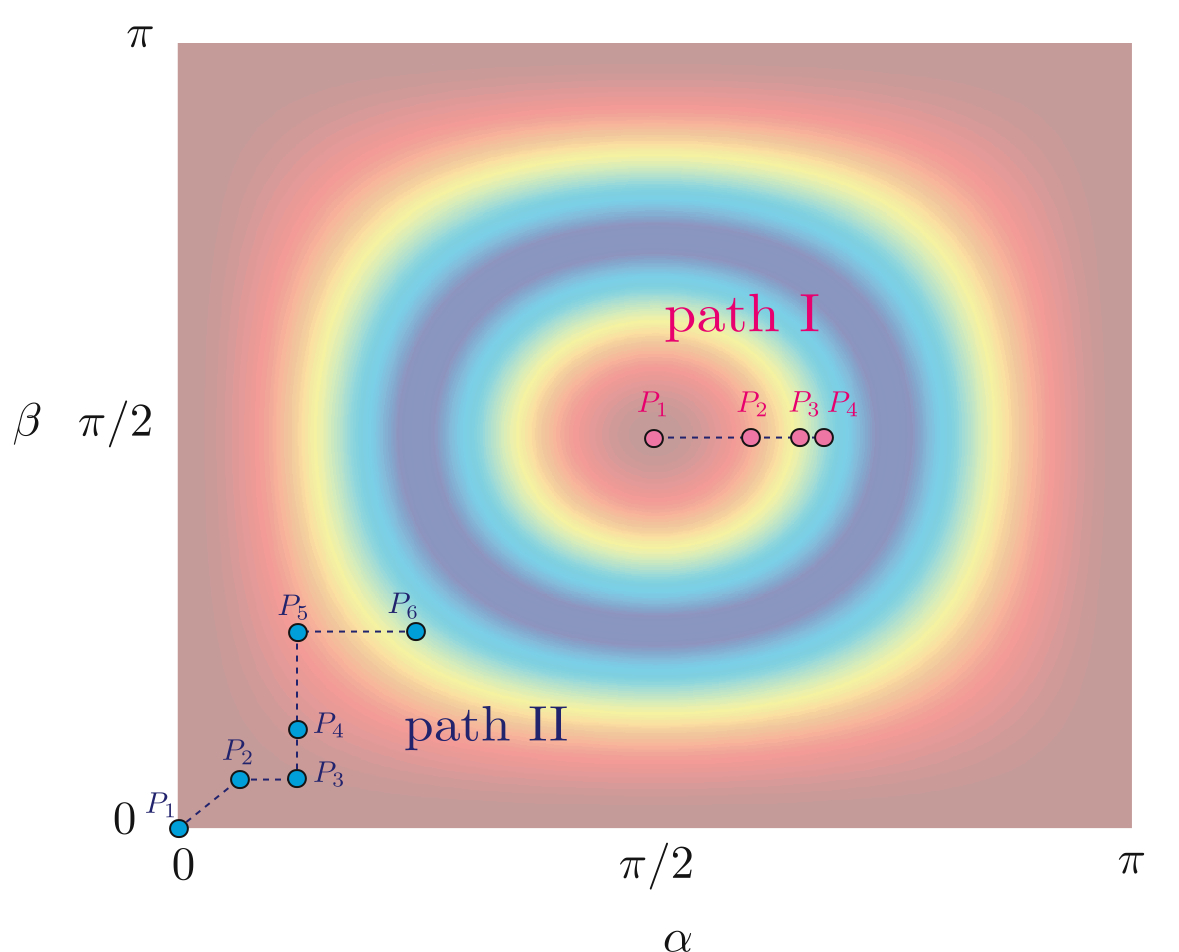}}}
\caption{\label{QHE4}  Two non-Abelian paths starting from the Graphene point (path I) and from the $\pi$-flux point (path II). The background contour corresponds to $\vert W(\alpha, \beta) \vert$} 
\end{center}
\end{figure} 
\end{center}

In this section, we consider the two different paths within the $\alpha-\beta$ space in Fig. \ref{QHE4}. The first path starts from the $\Pi$-flux point ($\alpha=\beta=\pi/2$) and reaches  $\alpha=0.9(3\pi/4) , \beta=\pi/2$ in four steps (see the purple dots in Fig. \ref{QHE4}). The second path starts from the graphene point ($\alpha=\beta=0$) and reaches  $\alpha=\beta=\pi/4$ in six steps (see the blue dots in Fig. \ref{QHE4}). Both paths teach us in what interesting manner the density profiles and the Hall plateaus are modified as the number of Dirac cones $N_d$ changes. \\

\begin{center} 
\begin{figure}
\begin{center}
\vspace{0.cm}{\scalebox{0.18}{\includegraphics{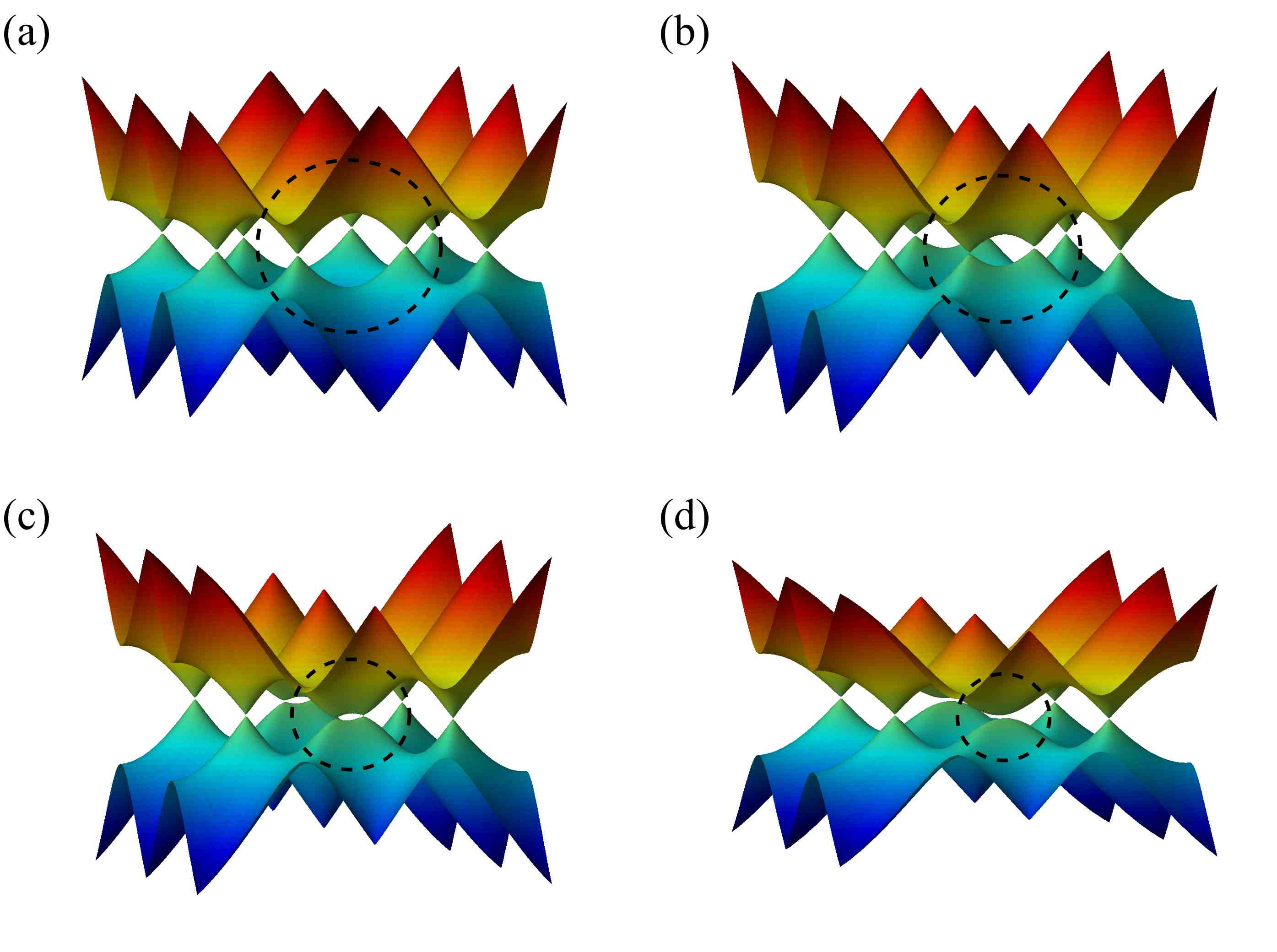}}}
\caption{\label{QHE8}  Energy spectra as one travels along the path I: (a) Point 1: $\alpha=\pi/2$, (b) Point 2:  $\alpha=3 \pi/5$,  (c)  Point 3:  $\alpha=0.87(3 \pi/4)$, (d)  Point 4: $\alpha=0.9(3\pi/4)$. The spectra are represented within the first BZ and one has set $\beta=\pi/2$ and $\Phi=0$. Dotted circles emphasize the merging of two central Dirac points.} 
\end{center}
\end{figure} 
\end{center}

\begin{center} 
\begin{figure}
\begin{center}
\vspace{0.cm}{\scalebox{0.25}{\includegraphics{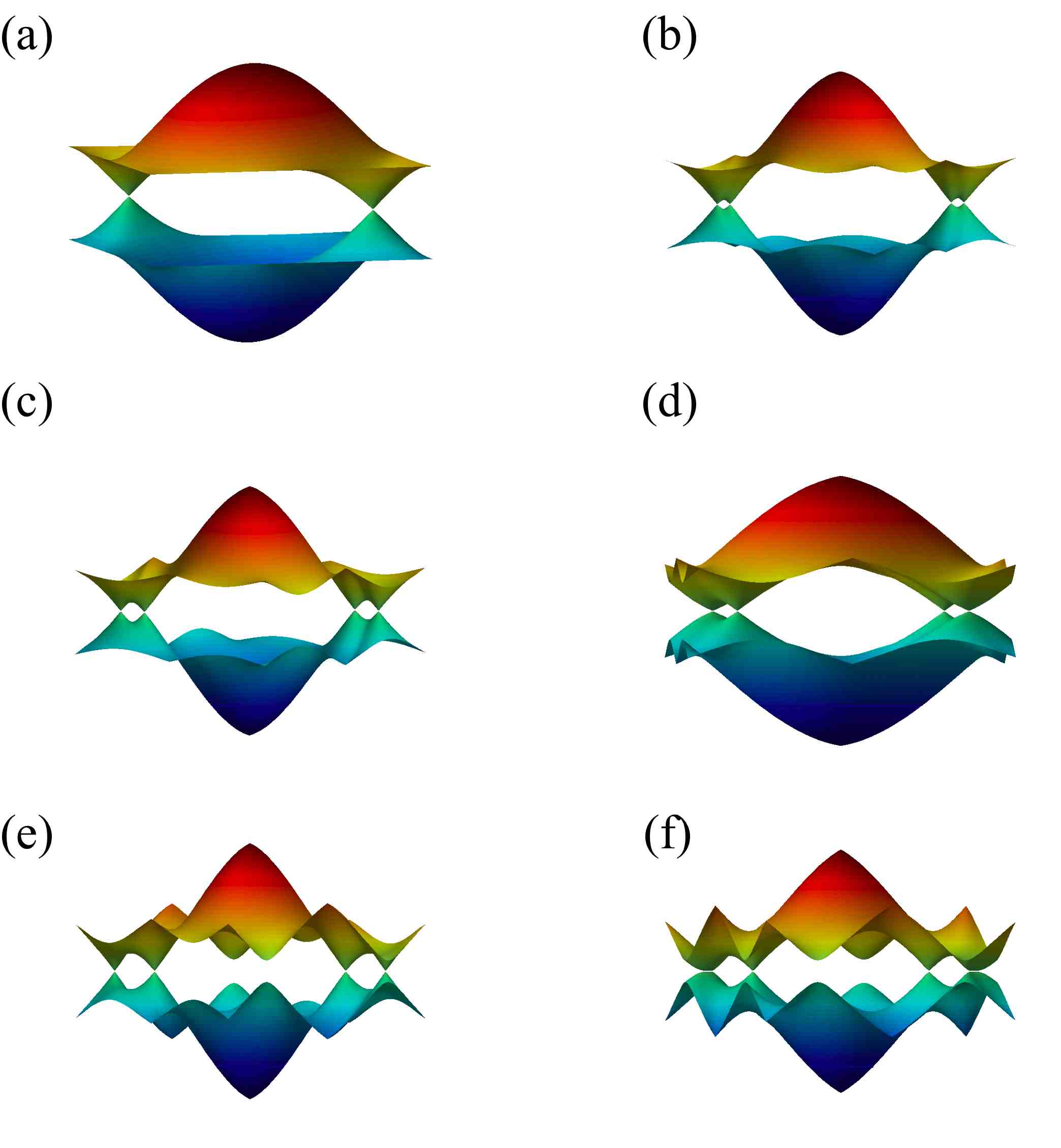}}}
\caption{\label{QHE5}  Energy spectra as one travels along the path II: (a) Point 1: $\alpha=\beta=0$, (b) Point 2:  $\alpha=\beta=\pi/16$,  (c)  Point 3: $\alpha=\pi/8, \beta=\pi/16$, (d)  Point 4: $\alpha=\beta=\pi/8$, (e)  Point 5: $\alpha=\pi/8, \beta=\pi/4$, (f)  Point 6: $\alpha= \beta=\pi/4$. The spectra are represented within the first BZ and one sets $\Phi=0$.} 
\end{center}
\end{figure} 
\end{center}

\vspace{1ex}
{\bf Path I.-} Let us first study the path starting at the $\Pi$-flux point (see path I in Fig. \ref{QHE4}). In this path, the parameter $\beta=\pi/2$ is constant, while $\alpha$ is varied from $\pi/2$ to $0.9(3\pi/4)$. The corresponding energy spectra at vanishing flux $\Phi=0$ are represented in Fig. \ref{QHE8}, where one observes how the eight initial isotropic Dirac points are progressively deformed as the flux is varied. As one travels between the Point 2 ($\alpha=3 \pi/5$) and the Point 3 ( $\alpha=0.87(3 \pi/4)$), one clearly observes that four Dirac points approach and finally merge at  Point 4 $\alpha=0.9(3\pi/4)$ (see Fig. \ref{QHE8} (b) and (c)). Indeed, at criticality $\alpha_c=3\pi/4$, the massless fermions merge into a highly anisotropic excitation, the latter being linear in one direction and parabolic in the transverse.
Below,  we explain how this topological phase transition is detected through density measurements and the AQHE.  Note that we  focus on very low fluxes $\Phi$, which allows one to discuss the Hall plateaus in terms of the energy spectra in Fig. \ref{QHE5}.

\begin{itemize}

\item Point 1: $\alpha=\pi/2$. We first compute $\sigma_H$ through the density profiles at the $\Pi$-flux point. The van Hove singularities (VHS) in the density of states (DOS)  delimit the relativistic regime between $E_{VHS}=\pm 0.4$, where we observe steps of eight in the Hall conductivity  $\sigma_{H}=\pm\frac{8}{h}(\nu+\frac{1}{2})$. This numerical computation  emphasizes the presence of the $N_d=8$ massless Dirac fermions, as observed in Fig. \ref{QHE8} (a).

\item Point 2 to Point 4: $\alpha=3 \pi/5 \rightarrow 0.9(3\pi/4)$. As the non-Abelian flux $\alpha$ is varied, we observe a transition from $\sigma_{H}=\pm\frac{8}{h}(\nu+\frac{1}{2})$ to $\sigma_{H}=\pm\frac{4}{h}(\nu+\frac{1}{2})$  around $E=0$. Consequently, the density profile measurements performed along this path witness the topological phase transition characterized by the change $N_d= 8 \rightarrow 4$.

\end{itemize}

The above analysis confirms the aforementioned conjecture announced in Eq. \eqref{conjecture}, and thus proves that density measurements do indeed witness the topological phase transition.

\vspace{1ex}
{\bf Path II.-} Let us then focus on the second path (see path II in Fig. \ref{QHE4}). The corresponding energy spectra at vanishing flux $\Phi=0$ are represented in Fig. \ref{QHE5}, where one observes that the two initial 2-degenerate Dirac points are progressively split into four non-degenerate singularities, and at some point  become anisotropic (see Fig. \ref{QHE5} (d) and (f)). We study below how these modifications affect the anomalous QHE. \\

\begin{itemize}

\item Point 1, $\alpha=\beta=0$: the graphene point has already been studied in the previous section, where the anomalous steps of four have been observed between $E \in [-1 , 1]$. This effect is due to the presence of two distinct 2-degenerate Dirac fermions $N_d=4$ at $\Phi=0$.

\item Point 2,  $\alpha=\beta=\pi/16$: as the degeneracy is lifted, four (non-degenerate) Dirac cones are present in the Brillouin zone. Our numerical results show that the Hall plateaus exhibit steps of four $\sigma_{H}=\pm\frac{4}{h}(\nu+\frac{1}{2})$ in the range $E \in [-0.11 , 0.11]$, steps of two   in the range $E \in [-1, -0.11)\bigcup(0.11 , 1]$, and steps of one elsewhere $|E|>1$. These three regimes are delimited by VHS in the DOS at $E_{\txt{VHS}^{1}}= \pm 0.11$, and at $E_{\txt{VHS}^{2}}= \pm 1$. This analysis suggests the existence of three different regimes, namely, a \emph{relativistic regime} with four isotropic Dirac cones, an \emph{intermediate relativistic regime} that is characterized by the reminiscent of the two initial cones, and finally a \emph{non-relativistic regime} where the linear dispersion vanishes. Additionally, we have numerically confirmed that the central plateau at $E_F=0$ (i.e. $\sigma_H=-2 \rightarrow 2$) is robust, which is in accordance with symmetry and topological considerations on the corresponding zero-mode~\cite{jackiw_qhe}.  Therefore,  the zero-energy Hall conductivity is an ideal witness of the number of emerging massless fermions.

\item Point 3 to Point 6, $\alpha=\pi/8, \beta=\pi/16 \rightarrow \alpha= \beta=\pi/4$: along this path, one observes the interesting fact that the density profiles, and thus the Hall plateaus, are affected by the anisotropic character of the Dirac cones. Hence, unusual behavior is observed in the \emph{relativistic regime} where the conductivity evolves with steps of two in the anisotropic regime (cf. Figs. \ref{QHE5}(d) and (f)), whereas it evolves with steps of four in the isotropic regime (cf. Figs. \ref{QHE5}(c) and (e)). One may conclude that the anisotropic cones play a peculiar role in the anomalous quantum Hall effect: these anisotropic singularities do not contribute to the Hall conductivity for $E_F \neq 0$.

\end{itemize}

\section{Experimental characterization of topological phases}
\label{detection}

In the previous section, we have described in detail a method of detection of the phenomena predicted in this paper using density profile measurements. This is the easiest and the most convenient method, and with current experimental setups can be realized with the accuracy of, say, 5 percent. It is reasonable  to expect that by optimizing this method, accuracies  of 1 percent are achievable.  However, since the detection in quantum many-body systems is one of the most challenging tasks,  we discuss in this section other possible measurements strategies to detect topological phases and Dirac fermions. We present the state-of-art of all methods and discuss their precision. All  the 
detection methods discussed below have at least some of the following  properties:

\begin{itemize}
\item {\it Spatial resolution.} That means that the methods allow to resolve spatial (or equivalently Fourier transformed momentum) characteristics of the system.
\item {\it Color resolution.} Note that the internal degrees of freedom used for decoding the color states will correspond to internal spin, or pseudo-spin states. Many methods will allow to resolve color states spectroscopically.
\item {\it Non-demolition property.} That means that the methods will affect the measured quantum mechanical systems in a ``minimalÕÕ fashion'', and can be even repeated to allow for time resolution. Only few methods have this property, such as contrast imaging of atomic density and spin polarization spectroscopy (see below). 
\end{itemize}

Detection methods can be divided into two groups: those which address directly the properties of Dirac fermions, and those that have more general character and provide general properties of the system. 

\vspace{1ex}

\paragraph{\bf Direct detection of Dirac fermions}
These methods can be further classified as: i) global detection mechanisms, that allow us to infer the total number of massless fermions (this is the reason for calling them global), and ii) local detection mechanism, that should allow us to map the whole Fermi surface, and thus locate each of the conical singularities.

Among the global detection mechanisms we can list:
\begin{itemize}

\item Measurement of the atom density as a function of the chemical potential: Following~\cite{duan_graphene_ol}, a measurement of the atom density close to zero chemical potential depends on the number of Dirac points in the Fermi surface. In Ref. \cite{duan_graphene_ol} it was proposed to detect massless-to-massive Dirac particles transition; here we propose to use it to detect the change of number of the Dirac fermions. In principle, this method requires nothing but a precise measurement of the density, as in the case of the method discussed in the previous section. The seminal steps toward the precise density measurements were achieved quite recently in the first observations of the wedding cake structure of the Mott insulator -superfluid transition in  the optical lattice loosely confined in an harmonic trap \cite{wedding_ketterle,wedding_bloch}. Perhaps the most recent state-of-art paper is Ref. \cite{validation} in which validation of the quantum simulator of MI-SF transition is realized. Here, the most accurate Quantum Monte Carlo simulations are directly compared with experimental data; in this way accuracies of 1 percent in density determination are reached. 

\end{itemize}

The local detection mechanisms are obviously more demanding; among them we can point out: 

\begin{itemize}
\item Atomic angle-resolved photoemission spectroscopy ARPES method: a possible candidate for the adaptation of ARPES techniques to optical lattice setups, is based on using Raman transitions to a side level that is decoupled from the system~\cite{arpes_ol}. This is a very powerful method that allows to determine directly the spatially resolved single particle density matrix, i.e. indirectly to map the Fermi surface. This new measurement technique has been realized for ultracold atom gases in  Ref. \cite{steward}; like photoemission spectroscopy for electronic materials, it directly probes low energy excitations and thus can reveal excitation gaps and/or pseudogaps with accuracy less than 5 percent. Atomic ARPES is a destructive method, but in principle allows for color resolution (by splitting the relevant states in an external fields and using the spectral resolution), i.e. it allows to measure $ \rho_{st}(x,xÕ)= \langle \Psi_s{\dag}(x)\Psi_t(xÕ)\rangle$, where $s,t$ are the color indices.

\item Direct band mapping  is an alternative approach \cite{band1,band2,band3}. In this method one releases adiabatically the optical potentials to measure directly the quasi-momentum distribution. This can be combined with contrast imaging, so that can be combined with other measurements, such as density measurement in the same experimental run (see for example \cite{mott_fermion_2}). This method allows to map higher bands independently, which in principle would determine positions of the Dirac points directly. Also, it can incorporate the  color resolution. 
\end{itemize}

\paragraph{\bf Indirect detection of Dirac fermions}
Apart from these direct methods, there is a whole variety of general detection methods that will provide additional information and characteristics of the system. Although such measurements will not give information about the number of Dirac points, they will characterize additional interesting aspects of the system:

\begin{itemize}

\item Density imaging: Of course, the density measurements are the necessary ingredient of the direct methods based on Streda formula, or dependence on chemical potential, discussed above. Here, we consider them in general sense as a source of additional information.  The density imaging  allows for color resolution, and can be done destructively using absorptive imaging, or non-destructively using contrast imaging. It is used in the band mapping technique, but can be also used if we turn the optical potentials suddenly off to measure the momentum distribution, which can give a particularly interesting characterization of the considered states of the system.

\item Noise interferometry: this very powerful (although destructive) method, proposed in Ref. \cite{altman_noise}, can be used to measure directly the two-point density-density correlation functions, which can be expressed in terms of  Fourier transforms in the (quasi) momentum space of  occupations of the corresponding single particle states. Originally applied to bosons \cite{bloch-noise1}, it is very useful to characterize fermionic states \cite{bloch-noise2}. It gives indirect information about density of states close to the Fermi energy.

\item Hall conductivity measurements: this is discussed in detail in Sec.~ \ref{QHEsect}, where we provide very detailed examples for predicted results of such measurements. This method   is in the first place an  indirect method. Still, since the  anomalous quantum Hall effect is highly affected by the Fermi surface topology at $E_F=0$, the Hall conductivity,  determined from density profiles measurements as in Refs. \cite{Umu,fuxiang,shao,fuxiang2}, provides direct information about  the change in the number of Dirac fermions.

\item Spin polarization spectroscopy: This method (see \cite{Eckert} and references therein) may be applied to measure Fourier component of the spin density and their fluctuations. The methods is based on the quantum Faraday effect, i.e.  coupling the polarized light beam to atoms, mapping the state of atoms onto light, and then ``reading" off the information about atoms by measuring the fluctuations of the light beam using homodyne detection.
It combines color and spatial resolution with non-demolition character, and is thus very promising and powerful. This method is very well developed experimentally (for a review see \cite{hammerer}), and the experiments with spatial, or, better to say, momentum resolution have been initiated in the group of E. Polzik.  Similarly, as noise interferometry, it gives indirect information about density of states close to the Fermi energy.

\item Momentum-resolved Bragg spectroscopy: this method was pioneered by W.D. Phillips \cite{kozuma} and W. Ketterle \cite{stenger,stamper-kurn} groups. It allows to measure the structure factor, spin structure factor and other momentum and frequency resolved correlation functions. It also gives direct information of the dispersion relation of  low-energy (quasi) particle excitations. The current state-of-art of this method is presented in Ref. \cite{Ernst}, where the accuracies of few percents are reached.

\item Methods involving cavity QED: these ideas were mostly developed by H. Ritsch and co-workers (\cite{mekhov1,mekhov2}, see also \cite{ye}), but have not yet been realized experimentally. They are somewhat similar to ARPES, with cavity-enhancing of the signal. In principle they are aimed on measuring the spatial density-density correlations by coupling the systems to the cavity mode, and measuring statistics of the light outgoing from the cavity.

\end{itemize}
Last, but not least, we should mention enormous progress in detecting atomic density in situ and in expanding clouds of a single atom resolution ~\cite{greiner_single_atom_resolution,bloch_private, schmiedmayer_single_atom_resolution}. These methods will add on unprecedented precision to many of the above discussed detection schemes.

\section{Conclusions}
\label{conclusions}

In this article we have discussed the optical-lattice analogue of graphene subjected to additional non-Abelian gauge fields. We have shown how a long-wavelength perspective captures the relativistic nature of the excitations responsible for transport properties of the system. Besides, when the effective fields are switched on, the topology of the Fermi surface is modified, and novel phases with gapless fermionic excitations that differ from graphene arise. Such topological phases have been fully characterized by a topological charge assigned to each of the massless fermions. 

We have also discussed how an adiabatic variation of the external fluxes along a non-Abelian path can be interpreted as a topological phase transition between different quantum orders. We have identified two processes responsible for such phase transitions, namely, spontaneous and stimulated scattering events where the number of gapless fermions is modified. 

We have demonstrated that the creation and annihilation processes not only modify the Fermi surface's topology but also lead to interesting generalizations of the anomalous QHE. Moreover, we proved that a measure of this effect through atomic density profiles gives a direct signature of the number of Dirac fermions. In this sense, we emphasized that the diverse topological phases are characterized by topological charges (winding numbers), but also by a specific sequence of the underlying Chern numbers.

These systems provide a very challenging, but at the same time rich and interesting playground for applying various known detection techniques. Some of them provide direct information about Dirac fermions and topological phase transition, the other provide additional, but highly relevant characteristics.    
Despite the fact that quite a lot of methods exist, it is desirable to search for new ones especially suited for the systems  considered here (attempting for instance to measure topological invariants directly). 

We discussed here the physics of single particles. Obviously, it opens an avenue towards studies of even more challenging physics of interacting ultracold atoms in non-Abelian gauge fields, such as the fractional QHE or the Fermi color-superconductivity.

Summarizing, the effect of non-Abelian gauge fields on the  emerging relativistic theory is far from trivial. Not only do such fields transport the topological charges attached to these Dirac points, but they also lead to scattering processes where pairs of fermions with opposite topological charges  are created or annihilated, as a result of the transition between different quantum orders.  The possibility to observe such exotic phenomena, and extensions thereof, in an optical lattice table-top experiment  seems to be a challenging and rewarding task where interdisciplinary effects of  condensed-matter, high-energy, and atomic physics can be observed.

\vspace{1ex}
{\it Acknowledgements.-}  A.B. and M.A.M.D acknowledge financial support from the projects FIS2006-04885, 
CAM-UCM/910758, and INSTANS 2005-2010. A. B. acknowledges support from a FPU MEC grant.A.K. and M.L. acknowledge Spanish MEC projects TOQATA (FIS2008-00784) and QOIT (Consolider Ingenio 2010), ESF/MEC project FERMIX (FIS2007-29996-E), EU Integrated Project  SCALA, EU STREP project NAMEQUAM, ERC Advanced Grant QUAGATUA. M.L. acknowledges also  Humboldt Foundation Senior Research Prize.
 A.K. acknowledges the Polish Government Scientific Funds
2009-2010. N.G. is financially supported by the F.R.S-F.N.R.S and he acknowledges P. Gaspard  and P. de Buyl for their encouragements and for technical support. The authors also thank I. Bloch, F. Guinea, and J. Dalibard for useful discussions.



\begin{references}

\bibitem{q_simulation}
R. Feynman, Int. J. Theo. Phys. {\bf 21,} 467 (1982).

\bibitem{qs_cond_mat_OL1}
M. Lewenstein, A. Sanpera, V. Ahufinger, B. Damski, A. Sen (De), and U. Sen, Adv. Phys. {\bf 56,} 243 (2007).


\bibitem{qs_cond_mat_OL2}
I. Bloch, J. Dalibard, and W. Zwerger, Rev. Mod. Phys. {\bf 80,} 885 (2008).

\bibitem{mott_greiner}
M. Greiner, O. Mandel, T. Esslinger, T.W. Hansch, and  I. Bloch, Nature {\bf 415,} 39 (2002).

\bibitem{mott_fermion_1}
R. J\"ordens, N. Strohmaier, K. G\"unter, H. Moritz, and T. Esslinger, Nature {\bf 455,} 204 (2008).

\bibitem{mott_fermion_2}
U. Schneider, L. Hackerm\"uller, S. Will, Th. Best, I. Bloch, T. A. Costi, R. W. Helmes, D. Rasch, and A. Rosch, Science {\bf 322,} 1520 (2008).



\bibitem{lewenstein_non_Abelian}
K. Osterloh, M. Baig, L. Santos, P. Zoller, and M. Lewenstein, Phys. Rev. Lett. \textbf{95}, 010403 (2005).

\bibitem{spielman} Several groups have started experiments in this direction; for one of the most promising approaches see Y.-J. Lin, W.D. Phillips, J.V. Porto, and I.B. Spielman, Bull. Am. Phys. Soc. {\bf 53}, No. 2, A14.00001 (2008); Y.-J. Lin, R.L. Compton, A.R. Perry, W.D. Phillips, J.V. Porto, and I.B. Spielman, Phys. Rev. Lett. { \bf 102, }130401 (2009); Y.ÐJ.  Lin, R.L. Compton, K. Jimenez-Garcia, J.V. Porto and I.B. Spielman, Nature {\bf 462,}  628 (2009).


\bibitem{duan_graphene_ol}
S.-L. Zhu, B. Wang, and L.-M. Duan, Phys. Rev. Lett. \textbf{98}, 260402 (2007).

\bibitem{graphene_exp}
K.S. Novoselov, A.K. Geim, S.V. Morozov, D. Jiang, Y. Zhang, S.V. Dubonos, I.V. Grigorieva, and A.A. Firsov , Science {\bf 306,} 666 (2004).


\bibitem{wen_book}
X. Wen, {\it Quantum Field Theory of Many-body Systems} (Oxford Univ. Press, Oxford, 2004).


\bibitem{graphite}
P. Wallace, Phys. Rev. {\bf 71,} 622 (1947).

\bibitem{semenoff}
G.W. Semenoff, Phys. Rev. Lett. \textbf{53}, 2449 (1984).

\bibitem{graphene_review}
A.H. Castro Neto, F. Guinea, N.M.R. Peres, K.S. Novoselov, and A.K. Geim, Rev. Mod. Phys. {\bf 81,} 109 (2009).

\bibitem{grap_klein_theo}
M.I. Katsnelson, K.S. Novoselov, and A.K. Geim, Nat. Phys. {\bf 2,} 620  (2006).

\bibitem{grap_klein_exp}
N. Stander, B. Huard, and D. Goldhaber-Gordon, Phys. Rev. Lett. {\bf 102,} 026807 (2009).
  




\bibitem{rel_landau_levels}
I.I. Rabi, Zeit. f. Physik {\bf 49,} 507 (1928)
 
 \bibitem{rel_landau_levels_graph}
 J.W.  McClure, Phys. Rev. {\bf 104,} 666 (1956).
 
 \bibitem{grap_landau_levels_exp}
G. Li and  E.Y. Andrei, Nat. Phys. {\bf 3,} 623  (2007).

\bibitem{cat_states}
A. Bermudez, M.A. Martin-Delgado, and E. Solano, Phys. Rev. Lett {\bf 99,} 123602 (2007).

\bibitem{jackiw_qhe}
R. Jackiw, Phys. Rev. D {\bf 29,} 2375 (1984).

\bibitem{schakel_qhe}
A.M.J. Schakel, Phys. Rev. D {\bf 43,} 1428 (1991).

\bibitem{sharapov_qhe}
V.P. Gusynin and S.G. Sharapov, Phys. Rev. Lett. \textbf{95}, 146801 (2005).

\bibitem{guinea_qhe}
N.M.R. Peres,  F. Guinea, and A.H. Castro Neto, 
Phys. Rev. B {\bf 73,} 125411(2006).


\bibitem{grap_anomalous_qhe_exp1}
K.S. Novoselov, A.K. Geim, S.V. Morozov, D. Jiang, M.I. Katsnelson, I.V. Grigorieva, S.V. Dubonos, and A.A. Firsov, Nature {\bf 438,} 197 (2005).

\bibitem{grap_anomalous_qhe_exp2}
Y. Zhang, Y.-W. Tan, H.L. Stormer, and  P. Kim, Nature {\bf 438,} 201 (2005).

\bibitem{bercioux_rombic}
D. Bercioux, D.F. Urban, H. Grabert, and W. H\"ausler, Phys. Rev. A {\bf 80,} 063603 (2009)








\bibitem{OL_Abelian_field}
D. Jaksch and P. Zoller, New J. Phys. {\bf 5,} 56 (2003).

\bibitem{OL_Abelian_field_2}
E.J. Mueller, Phys. Rev. A {\bf 70,} 041603(R) (2004).

\bibitem{OL_Abelian_field_3}
A.S. S{\o}rensen,  E. Demler, and M.D. Lukin, Phys. Rev. Lett. \textbf{94}, 086803 (2004).

\bibitem{OL_ab_field_dark_states_1}
G. Juzeli\=unas, P. \"Ohberg, J. Ruseckas, and A. Klein, Phys. Rev. A \textbf{71,} 053614 (2005).

\bibitem{OL_ab_field_dark_states_2}
G. Juzeli\=unas, and P. \"Ohberg, Phys. Rev. Lett. {\bf 93,} 033602 (2004).



\bibitem{polini}
M. Polini, R. Fazio, M.P. Tosi, J. Sinova, and A.H. MacDonald, Laser Physics  {\bf 14,} 603 (2004). 
	
\bibitem{polini2}
M. Polini, R. Fazio, A.H. MacDonald, and M.P. Tosi,  Phys. Rev. Lett. {\bf 95,} 010401 (2005).

\bibitem{Holland1}R. Bhat, L.D. Carr, and M.J. Holland, Phys. Rev. Lett. {\bf96}, 060405 (2006).

\bibitem{Holland2}R. Bhat, M. Kr\"amer, J. Cooper, and M.J. Holland,  Phys. Rev. A {\bf76}, 043601 (2007).

\bibitem{Tung} S. Tung, V. Schweikhard, and E.A. Cornell, Phys. Rev. Lett. {\bf97}, 240402 (2006).

 \bibitem{imme} A. Klein and D. Jaksch, Europhys. Lett. {\bf 85}, 13001 (2009).



\bibitem{OL_iqhe}
N. Goldman and P. Gaspard, Europhys. Lett. {\bf 78,} 60001 (2007).

\bibitem{nathan-book} For an introduction to the subject and a survey of results see N. Goldman, \emph{Quantum transport in lattices subjected to external gauge fields}, VDM Verlag (2009).


\bibitem{anomalous_qhe_ol}
 L.B. Shao, S.-L. Zhu, L. Sheng, D.Y. Xing, and Z.D. Wang, Phys. Rev. Lett. {\bf 101,} 246810 (2008).
 
 \bibitem{anomalous_qhe_kondo}
X. Chen, S. Dong, and J.-M. Liu, arXiv:0909.3142 (2009).


\bibitem{OL_fqhe}
R.N. Palmer and D. Jaksch, Phys. Rev. Lett. {\bf 96,} 180407 (2006).




\bibitem{fleischhauer_non_ab}
J. Ruseckas, G. Juzeli\=unas, P. \"Ohberg, and M. Fleischhauer, Phys. Rev. Lett. \textbf{95}, 010404 (2005).

\bibitem{cavity_non_Abelian}
J. Larson and S. Levin, Phys. Rev. Lett.  {\bf 103,} 013602 (2009).

\bibitem{non_ab_spin_hall_effect}
S.-L. Zhu, H. Fu, C.-J. Wu, S.-C. Zhang, and L.-M. Duan, Phys. Rev. Lett. {\bf 97,} 240401 (2006).

\bibitem{non_ab_MI_transition}
I.I. Satija, D.C. Dakin, and C.W. Clark, Phys. Rev. Lett. {\bf 97,} 216401 (2006).

\bibitem{non_ab_quasi_relat}
G. Juzeli\=unas, J. Ruseckas, M. Lindberg, L. Santos, and P. \"Ohberg, Phys. Rev. A {\bf 77,} 011802(R) (2008). 

\bibitem{non_ab_zitterbewegung}
J.Y. Vaishnav and C.W. Clark, Phys. Rev. Lett. {\bf 100,} 153002 (2008).


\bibitem{non_ab_neg_refraction}
G. Juzeli\=unas, J. Ruseckas, A. Jacob, L. Santos, and P. \"Ohberg, Phys. Rev. Lett. {\bf 100,} 200405 (2008).

\bibitem{non_ab_spin_transisitor}
J.Y. Vaishnav, J. Ruseckas, C.W. Clark, and G. Juzeli\=unas, Phys. Rev. Lett. {\bf 101,} 265302 (2008).

\bibitem{non_ab_non_localization}
S.-L. Zhu, D.-W. Zhang, and Z.D. Wang, Phys. Rev. Lett. {\bf 102,} 210403 (2009).

\bibitem{non_ab_monopole}
V. Pietil\"a and M. M\"ott\"onen, Phys. Rev. Lett. {\bf 102,} 080403 (2009).

\bibitem{pachos}
P. Maraner and J.K. Pachos, Phys. Lett. A {\bf 373,} 2542 (2009).

\bibitem{so_coupling}
X.-J. Liu, M.F. Borunda, X. Liu, and J. Sinova, Phys. Rev. Lett. {\bf 102,} 046402 (2009).

\bibitem{top_order_p_ip_sf}
C. Zhang, S. Tewari, R. M. Lutchyn, and S. Das Sarma, Phys. Rev. Lett. {\bf 101,} 160401 (2008).

\bibitem{top_order_p_ip_sf_2}
M. Sato, Y. Takahashi, and S. Fujimoto, Phys. Rev. Lett. {\bf 103}, 020401 (2009). 



\bibitem{OL_non_ab_qhe}
N. Goldman, A. Kubasiak, P. Gaspard, and M. Lewenstein, Phys. Rev. A. {\bf 79,} 023624 (2009).

\bibitem{dirac_fermions_non_ab_square}
N. Goldman, A. Kubasiak, A. Bermudez, P. Gaspard,  M. Lewenstein, and M.A. Martin Delgado, 
Phys. Rev. Lett. {\bf 103}, 035301 (2009).



\bibitem{dirac_fermions_square_staggered_1}
 L.-K. Lim, C.M. Smith, and A. Hemmerich, Phys. Rev. Lett. {\bf 100,} 130402 (2008).

 
\bibitem{dirac_fermions_square_staggered_2}
J.-M. Hou, W.-X. Yang, and X.-J. Liu, Phys. Rev. A {\bf 79,} 043621 (2009).


\bibitem{wen_zee}
X. Wen and A. Zee, Nucl. Phys. B {\bf 316,} 641 (1989).

\bibitem{dirac_fermions_OL}
K.L. Lee, B. Gremaud, R. Han, B.-G. Englert, and C. Miniatura,  Phys. Rev. A {\bf  80,} 043411 (2009).

\bibitem{sengstock} see http://www.physnet.uni-hamburg.de/ilp/sengstock/index.html; A. Eckardt, Ph. Hauke, P. Soltan-Panahi, C. Becker, K. Sengstock, and M. Lewenstein, 
arXiv:0907.0423.



\bibitem{feschbach}
E. Timmermans, P. Tommasini, M. Hussein, and A. Kerman, Phys. Rep. {\bf 315,} 199 (1999).


\bibitem{kekule_graphene_1}
C.-Y. Hou, C. Chamon, and C. Mudry, Phys. Rev. Lett. {\bf 98,} 186809 (2007).

\bibitem{kekule_graphene_2}
R. Jackiw and S.-Y. Pi, Phys. Rev. Lett. {\bf 98,} 266402 (2007).


\bibitem{volovik}
G.E. Volovik, Lect. Notes in Phys.  {\bf 718,} 31-73 (2007).

\bibitem{wen_fqhe}
X. Wen, Adv. Phys. {\bf 44,} 405 (1995).

\bibitem{wen_spin_liquid}
X. Wen, Phys. Rev B {\bf 65,} 165113 (2002).

\bibitem{wen_zee_spin_liquid}
X. Wen and A. Zee, Phys. Rev B {\bf 66,} 235110 (2002).





 \bibitem{qshi_experiment}
 M. Koenig, S. Wiedmann, C. Br\"une, A. Roth, H. Buhmann, L.W. Molenkamp, X.-L. Qi, and S.-C. Zhang, { Science} \textbf{318,} 766 (2007). 
 
  \bibitem{3d_ti_experiment}
D. Hsieh, D. Qian, L. Wray, Y. Xia, Y.S. Hor, R. J. Cava, M.Z. Hasan, {Nature}  \textbf{452,} 970 (2008).
  
 \bibitem{3d_ti_experiment2}
D. Hsieh,Y. Xia, L. Wray, D. Qian, A. Pal, J. H. Dil, J. Osterwalder, F. Meier, G. Bihlmayer, C.L. Kane, Y. S. Hor, R.J. Cava, and M.Z. Hasan, Science  {\bf 323,} 919 (2009).
 
\bibitem{top_insulator_nv}
C.L. Kane and E.J.  Mele, {Science} {\bf 314,} 1692 (2006).
 
\bibitem{qsh_insulator_graphene}
C.L. Kane and E.J.  Mele, {Phys. Rev. Lett.}  \textbf{95,} 146802 (2005).


\bibitem{qsh_insulator_z2}
C.L. Kane and E.J.  Mele, {Phys. Rev. Lett.}  \textbf{95,} 226801 (2005).

  \bibitem{top_insulator_2}
 R. Roy, Phys. Rev. B {\bf 79,} 195321 (2009).

\bibitem{qshi_semicond}
B.A. Bernevig, T.L. Hughes, and  S.-C. Zhang, { Science} \textbf{314,} 1757 (2006). 
 
\bibitem{3d_top_insulator}
  L.Fu,  C.L. Kane, and E.J. Mele, { Phys. Rev. Lett.}  \textbf{98,} 106803 (2007).
  

   
  \bibitem{3d_top_insulator_3}
J.E. Moore and L. Balents, { Phys. Rev. B}  \textbf{75,} 121306(R) (2007).
  
  \bibitem{graphene_DP_merging}
Y. Hasegawa, R. Konno, H. Nakano, and M. Kohmoto, Phys. Rev. B {\bf 74,} 033413 (2006). 

\bibitem{garphene_DP_merging_2}
P. Dietl, F. Piechon, and G. Montambaux, Phys. Rev. Lett. {\bf 100,} 236405 (2008).

\bibitem{graphene_DP_merging_strain}
V.M. Pereira, A.H. Castro Neto, and N.M.R. Peres,  Phys. Rev. B {\bf 80,} 045401 (2009).

\bibitem{DP_merging_OL}
B. Wunsch, F. Guinea, and F. Sols, New J. Phys. {\bf 10,} 103027 (2008). 

\bibitem{DP_semi}
S. Banerjee, R.R.P. Singh, V. Pardo, and W.E.Picket, Phys. Rev. Lett. {\bf 103,} 016402 (2009).

\bibitem{DP_kz}
A. Dutta, R.R.P. Singh, and U. Divakaran, arXiv:0910.3896 (2009). 
 
\bibitem{Umu} R.O. Umucalilar, H. Zhai, and M. \"Oktel, Phys. Rev. Lett. {\bf 100}, 070402 (2008).

 \bibitem{hofstadter}D. Hofstadter, Phys. Rev. B {\bf 14}, 2239 (1976).
  
   \bibitem{Kohmoto1985} M. Kohmoto, Ann. Phys. {\bf 160}, 343 (1989).
 
 \bibitem{Fukui}T. Fukui, Y. Hatsugai, and H. Suzuki, J. Phys. Soc. Jap. \textbf{74}, 1674 (2005).
 

 
 \bibitem{fuxiang} F. Li, L. Sheng, and D.Y. Xing, Europhys. Lett. {\bf 84}, 60004 (2008).
 
  \bibitem{shao} L.B. Shao, S.-L. Zhu, L. Sheng, D.Y. Xing, and Z.D. Wang, Phys. Rev. Lett. {\bf 101}, 246810 (2008).
  
   \bibitem{fuxiang2} F. Li, L.B. Shao, L. Sheng, and D.Y. Xing, Phys. Rev. A  {\bf 78},  053617 (2008).
 
\bibitem{wedding_bloch} S. F\"olling, A. Widera, T. M\"uller, F. Gerbier, and I. Bloch, Phys. Rev. Lett. {\bf 97,} 060403 (2006). 

\bibitem{wedding_ketterle} G.K. Campbell, J. Mun, M. Boyd, P. Medley, A.E. Leanhardt, L. Marcassa, D.E. Pritchard, and W.  Ketterle, Science {\bf 313,}   649-652 (2006). 

\bibitem{validation} S. Trotzky, L. Pollet, F. Gerbier, U. Schnorrberger, I. Bloch, N.V. Prokof'ev, B. Svistunov, and M. Troyer, arXiv:0905.4882.


\bibitem{arpes_ol}
T.-L. Dao, A. Georges, J. Dalibard, C. Salomon, and I. Carusotto, Phys. Rev. Lett. {\bf 98,} 240402 (2007).

\bibitem{steward} J.T. Stewart, J.P. Gaebler, and D.S. Jin, Nature {\bf 454}, 744 (2008). 

\bibitem{band1} A. Kastberg, W.D. Phillips, S.L. Rolston, R.J.C. Spreeuw, and P.S. Jessen,  Phys. Rev. Lett. {\bf 74}, 1542 (1995).

\bibitem{band2} M. Greiner, I. Bloch, O. Mandel, T.W. Hansch, and T. Esslinger, Phys. Rev. Lett. {\bf 87}, 160405 (2001).

\bibitem{band3} M. K\"ohl, H. Moritz, T. St\"oferle, K. G\"unter, and T. Esslinger, Phys. Rev. Lett. {\bf 94}, 080403 (2005).

\bibitem{altman_noise} E. Altman, E. Demler, and M.D. Lukin, Phys. Rev. A {\bf 70}, 013603 (2004).



\bibitem{bloch-noise1} S. Foelling, F. Gerbier, A. Widera, O. Mandel, T. Gericke, and I. Bloch, Nature {\bf 434}, 481 (2005). 


\bibitem{bloch-noise2} T. Rom, Th. Best, D. van Oosten, U. Schneider, S. Foelling, B. Paredes, and I. Bloch,  Nature, {\bf 444}, 733 (2006). 


\bibitem{Eckert} K. Eckert, O. Romero-Isart, M. Rodriguez, M. Lewenstein, E.S. Polzik, and A. Sanpera, Nature Phys. {\bf 4}, 50 (2008). 

\bibitem{hammerer} K. Hammerer, A.. S{\o}rensen, and E.S. Polzik, arXiv:0807.3358, to appear in Rev. Mod. Phys. (2009).
 
 
 \bibitem{kozuma} M. Kozuma, L. Deng, E.W. Hagley, J. Wen, R. Lutwak, K. Helmerson, S.L. Rolston, and W.D. Phillips,  Phys. Rev. Lett. {\bf 82}, 871 (1999).


\bibitem{stenger} J. Stenger, S. Inouye, A.P. Chikkatur, D.M. Stamper-Kurn, D.E. Pritchard, and W. Ketterle,  Phys. Rev. Lett. {\bf 82}, 4569 (1999).


\bibitem{stamper-kurn}      D. M. Stamper-Kurn, A. P. Chikkatur, A. Gšrlitz, S. Inouye, S. Gupta, D. E. Pritchard, and W. Ketterle,  Phys. Rev. Lett. {\bf  83}, 2876 (1999).
                   
\bibitem{Ernst} P.T. Ernst, S. G{\"o}tze, J.S. Krauser, K. Pyka, D.-S. L{\"u}hmann, D. Pfannkuche, and K. Sengstock, 	Nature Phys. {\bf 6,} 56  (2009).



\bibitem{mekhov1} I.B. Mekhov, C. Maschler, and H. Ritsch, Phys. Rev. Lett. {\bf 98}, 100402 (2007). \bibitem{mekhov2} I.B. Mekhov, C. Maschler, and H. Ritsch, Nature
Phys. {\bf 3}, 319  (2007).

\bibitem{ye} J. Ye, J. Zhang, W. Liu, K. Zhang, Y. Li, W. Zhang, arXiv:0812.4077

 
 \bibitem{greiner_single_atom_resolution}W.S. Bakr, J.I. Gillen, A. Peng, S. Foelling, and M. Greiner,  Nature {\bf 462,} 7269 (2009).
 
 \bibitem{bloch_private}I. Bloch, private communication.
 
 \bibitem{schmiedmayer_single_atom_resolution}R. B\"ucker,  A. Perrin, S. Manz, T. Betz, Ch. Koller, T. Plisson, J. Rottmann, T. Schumm, and J. Schmiedmayer, New J. Phys. {\bf 11,} 103039 (2009).
 
















\end{references}
\end{document}